\documentclass[twocolumn]{aastex61}

\usepackage{amsmath}

\newcommand{\psr}{J1107--5907}
\newcommand{\mwaN}{86} 
\newcommand{\utmostN}{283} 
\newcommand{\simN}{51} 
\newcommand{\alphaMP}{-1.85 \pm 0.08} 
\newcommand{\alphaMPstd}{0.50} 
\newcommand{\alphaPR}{-2.21 \pm 0.10} 
\newcommand{\alphaPRstd}{0.35} 
\newcommand{\alphaPC}{-3.5} 
\newcommand{\DMobs}{40.75 \pm 0.02} 
\newcommand{\DMcat}{40.2 \pm 1.1} 
\newcommand{\DMoffset}{0.55} 
\newcommand{\RMobs}{23.85 \pm 0.08} 
\newcommand{\RMion}{-2.1 \pm 0.2} 
\newcommand{\RMism}{25.95 \pm 0.28} 

\newcommand{\Curtin}{International Centre for Radio Astronomy Research, Curtin University, Bentley, WA 6102, Australia}
\newcommand{\CAASTRO}{ARC Centre of Excellence for All-sky Astrophysics (CAASTRO), Australia}
\newcommand{\CSIRO}{CSIRO Astronomy and Space Science, P.O. Box 76, Epping, NSW 1710, Australia}
\newcommand{\ASU}{School of Earth and Space Exploration, Arizona State University, Tempe, AZ 85287, USA}
\newcommand{\USydney}{Sydney Institute for Astronomy, School of Physics, The University of Sydney, NSW 2006, Australia}
\newcommand{\ASTRO}{ARC Centre of Excellence for All Sky Astrophysics in 3 Dimensions (ASTRO 3D), Australia}
\newcommand{\UToronto}{Dunlap Institute for Astronomy and Astrophysics, University of Toronto, ON, M5S 3H4, Canada}
\newcommand{\UWM}{Department of Physics, University of Wisconsin--Milwaukee, Milwaukee, WI 53201, USA}
\newcommand{\UW}{Department of Physics, University of Washington, Seattle, WA 98195, USA}
\newcommand{\UWA}{International Centre for Radio Astronomy Research, University of Western Australia, Crawley, WA 6009, Australia}
\newcommand{\CSIROPerth}{CSIRO Astronomy and Space Science, P.O. Box 1130, Bentley, WA 6102, Australia}
\newcommand{\Swin}{Centre for Astrophysics and Supercomputing, Swinburne University of Technology, P.O. Box 218, Hawthorn, VIC 3122, Australia}
\newcommand{\Jodrell}{Jodrell Bank Centre for Astrophysics, School of Physics and Astronomy, The University of Manchester, Manchester M13 9PL, UK}
\newcommand{\ASTRON}{ASTRON, Netherlands Institute for Radio Astronomy, Oude Hoogeveensedijk 4, 7991 PD, Dwingeloo, The Netherlands}

\shorttitle{PSR \psr\ at low radio frequencies}
\shortauthors{Meyers et al.}

\begin{document}

\title{Hunting for radio emission from the intermittent pulsar \psr\ at low frequencies}

\correspondingauthor{B.~W.~Meyers}
\email{bradley.meyers@postgrad.curtin.edu.au}

\author[0000-0001-8845-1225]{B.~W.~Meyers}
\affiliation{\Curtin}
\affiliation{\CAASTRO}
\affiliation{\CSIRO}

\author[0000-0001-7662-2576]{S.~E.~Tremblay}
\affiliation{\Curtin}
\affiliation{\CAASTRO}

\author[0000-0002-8383-5059]{N.~D.~R.~Bhat}
\affiliation{\Curtin}
\affiliation{\CAASTRO}

\author[0000-0003-1110-0712]{C.~Flynn}
\affiliation{\Swin}
\affiliation{\CAASTRO}

\author[0000-0001-9817-4938]{V.~Gupta}
\affiliation{\Swin}

\author[0000-0002-7285-6348]{R.~M.~Shannon}
\affiliation{\Swin}

\author[0000-0003-3059-3823]{S.~G.~Murray}
\affiliation{\Curtin}
\affiliation{\CAASTRO}
\affiliation{\ASTRO}

\author[0000-0002-8950-7873]{C.~Sobey}
\affiliation{\CSIROPerth}

\author[0000-0002-6380-1425]{S.~M.~Ord}
\affiliation{\CSIRO}

\author[0000-0003-0289-0732]{S.~Os\l{}owski}
\affiliation{\Swin}

\author{B.~Crosse}
\affiliation{\Curtin}

\author{A.~Williams}
\affiliation{\Curtin}

\author[0000-0002-6658-2811]{F.~Jankowski}
\affiliation{\Jodrell}
\affiliation{\Swin}
\affiliation{\CAASTRO}

\author[0000-0002-0161-7243]{W.~Farah}
\affiliation{\Swin}

\author{V.~Venkatraman~Krishnan}
\affiliation{\Swin}
\affiliation{\CAASTRO}

\author{T.~Bateman}
\affiliation{\USydney}

\author{M.~Bailes}
\affiliation{\Swin}

\author[0000-0001-9428-8233]{A.~Beardsley}
\affiliation{\ASU}

\author[0000-0002-4058-1837]{D.~Emrich}
\affiliation{\Curtin}

\author[0000-0001-6449-9611]{T.~M.~O.~Franzen}
\affiliation{\ASTRON}

\author[0000-0002-3382-9558]{B.~M.~Gaensler}
\affiliation{\UToronto}

\author{L.~Horsley}
\affiliation{\Curtin}

\author[0000-0003-2756-8301]{M.~Johnston-Hollitt}
\affiliation{\Curtin}

\author[0000-0001-6295-2881]{D.~L.~Kaplan}
\affiliation{\UWM}

\author{D.~Kenney}
\affiliation{\Curtin}

\author[0000-0001-7694-4030]{M.~F.~Morales}
\affiliation{\UW}
\affiliation{\ASTRO}

\author{D.~Pallot}
\affiliation{\UWA}

\author{K.~Steele}
\affiliation{\Curtin}

\author[0000-0002-8195-7562]{S.~J.~Tingay}
\affiliation{\Curtin}

\author[0000-0001-6324-1766]{C.~M.~Trott}
\affiliation{\Curtin}
\affiliation{\ASTRO}

\author{M.~Walker}
\affiliation{\Curtin}

\author[0000-0002-6995-4131]{R.~B.~Wayth}
\affiliation{\Curtin}
\affiliation{\ASTRO}

\author{C.~Wu}
\affiliation{\UWA}

\begin{abstract}
Rare intermittent pulsars pose some of the most challenging questions surrounding the pulsar emission mechanism, but typically have relatively minimal low-frequency ($\lesssim 300$\,MHz) coverage.
We present the first low-frequency detection of the intermittent pulsar \psr\ with the Murchison Widefield Array (MWA) at 154\,MHz and the simultaneous detection from the recently upgraded Molonglo Observatory Synthesis Telescope (UTMOST) at 835\,MHz, as part of an on-going observing campaign.
During a 30 minute simultaneous observation, we detected the pulsar in its bright emission state for approximately 15 minutes, where \mwaN\ and \utmostN\ pulses were detected above a signal-to-noise threshold of 6 with the MWA and UTMOST, respectively.
Of the detected pulses, 51 had counterparts at both frequencies and exhibited steep spectral indices for both the bright main pulse component and the precursor component.
We find that the bright state pulse energy distribution is best parameterized by a log-normal distribution at both frequencies, contrary to previous results that suggested a power law distribution.
Further low-frequency observations are required in order to explore in detail aspects such as pulse-to-pulse variability and intensity modulations, as well as to better constrain the signal propagation effects due to the interstellar medium and intermittency characteristics at these frequencies.
The spectral index, extended profile emission covering a large fraction of pulse longitude, and the broadband intermittency of PSR \psr\ suggest that future low-frequency pulsar searches---for instance those planned with SKA-Low---will be in an excellent position to find and investigate new pulsars of this type.
\end{abstract}


\section{Introduction}\label{sec:intro}
Pulsars that only emit occasionally are not only difficult to detect, but pose a fundamental challenge to understanding pulsar emission physics.
The duty cycle and timescales vary drastically for the pulsar population subclasses and individual pulsars within them, from seconds to minutes (i.e. nulling, \citealt{1970NatureBackerA}), hours to days (i.e. Rotating Radio Transients, RRATs, \citealt{2006NatureMcLaughlin}), and in the most extreme cases, months or years (i.e. intermittent pulsars, \citealt{2006SciKramer}). 
How the pulsar magnetospheric configuration changes so severely as to halt emission on short timescales (suddenly within a few rotations of the neutron star) is unclear, even when excluding the added complexity of the wide variety of timescales on which these state-switching phenomena occur.

The geometry of a pulsar is thought to play a significant role in governing the salient properties of its radio emission.
It has been shown, for example, that there is a strong correlation between the ``intermittency'' of a pulsar and the alignment of the magnetic and spin axes \citep{2008ApJCordes}.
There are myriad models for sporadic radio emission from pulsars to explain phenomena such as nulling or RRATs, including: magnetospheric plasma properties, particle supply, and acceleration region evolution (e.g. \citealp{2010MNRASTimokhin,2012ApJLi,2014MNRASMelrose,2015MNRASSzary}), interstellar dust interactions with currents within the magnetosphere \citep{1985ApJCheng}, and mechanisms such as circumpulsar plasma disks \citep{1981ApJMichel} or asteroidal debris \citep{2008ApJCordes,2013A&AMottez}.
Others suggest that, in some systems, the intermittency is related to free precession \citep{2006MNRASAkgun,2012MNRASJones}, or is a chaotic process (e.g. \citealp{2013MNRASSeymour}).
Notably, sporadic emission from pulsars is also thought to be a broadband phenomenon based on contemporaneous high-energy and radio observations.
An example of this broadband nature is the observed mode changing (i.e. where the profile and pulsar spin properties change significantly and abruptly, \citealt{2010SciLyne}) of PSRs B0943+10 \citep{2013SciHermsen,2016ApJMereghetti} and B0823+26 \citep{2018MNRASHermsen} in contemporaneous radio and X-ray data.
There is no consensus regarding which model is favored, and testing these hypotheses is notoriously difficult---only now becoming possible with the next generation of space and ground-based telescopes.

There are five confirmed examples of intermittent pulsars in the literature: PSRs \psr\ \citep{2006MNRASLorimer,2006ChJASOBrien,2014MNRASYoung,2016MNRASHobbs}, J1717--4054 (B1713--40, \citealp{1992MNRASJohnston,2014MNRASKerr}), J1832+0029 \citep{2006MNRASLorimer,2012ApJLorimer}, J1841--0500 \citep{2012ApJCamilo}, and J1933+2421 (B1931+24, \citealt{1985NatStokes,2006SciKramer,2008MNRASRea,2013MNRASYoung}).
Most of these pulsars were discovered well before being classified as intermittent, implying that it is quite possible that there are other examples of this kind of sporadic behavior in the $>2600$ known pulsars in the ATNF Pulsar Catalogue\footnote{\url{http://www.atnf.csiro.au/research/pulsar/psrcat/}} \citep{2005AJManchester}.
It also follows that there could be many intermittent pulsars that have been missed in pulsar surveys, thus biasing the known pulsar populations.
Intermittent pulsars have been recognized as a population relatively recently, and thus are not particularly well-studied, other than perhaps the prototypical example of J1933+2421.
The physics governing the emission behavior of these pulsars is likely closely linked to the same mechanisms responsible for nulling, the RRAT phenomenon (e.g. \citealt{2014MNRASYoung}) and possibly mode-changing.
In all cases, the frequency coverage over which intermittent pulsars have been observed is relatively small and does not include any low-frequency information (i.e. $\lesssim 300$\,MHz).
Low frequencies are thought to probe substantially different regions of the pulsar magnetosphere, provide improved constraints on the shape of the pulsar emission spectrum, and allow us to sample frequency dependent intermittency rates (e.g. \citealp{2003ApJMcLaughlin,2009ApJDeneva}), particularly if observed regularly and simultaneously over a wide frequency range.

PSR \psr\ is a relatively old ($\approx 440$\,Myr) and isolated pulsar discovered in the Parkes Multibeam Pulsar Survey (PMPS; \citealp{2001MNRASManchester,2006MNRASLorimer}), with a period of $P\approx 0.253$\,s and a moderate cataloged dispersion measure of $\DMcat \mathrm{\,pc\,cm^{-3}}$.
It is the brightest and one of the most active (in terms of its ``off'' timescale being relatively small for frequencies $\geq 700$\,MHz) of the intermittent pulsar population (see, e.g., Figure 10 of \citealt{2016MNRASHobbs}).
The pulsar was investigated after its discovery by \citet{2006ChJASOBrien}, who identified three distinct emission states: a bright state with a wide profile, a weak state with a narrow profile, and an ``off'' state.
\citet{2014MNRASYoung} used Parkes monitoring data collected over a decade to further explore the properties of each of the emission states, finding that the bright state has a complex temporal structure, with a relatively wide main pulse and lower-level emission across a large fraction of the pulsar rotation.
The weak state exhibited almost exclusively single-pulse events (though these were analyzed by creating subintegrations) without any particular phase localization and showed an extreme level of nulling. 
The ``off'' state consists of no observable emission, though this can be difficult to disentangle from the weak emission state.
Polarization analysis of the Parkes data also revealed that PSR \psr\ is consistent with being a nearly aligned rotator, which is reinforced by both its broad emission profile and large characteristic age \citep{1990ApJRankin,1998MNRASTauris,2008MNRASWeltevrede,2010MNRASYoung}.

Typically, PSR \psr\ switches between bright and weak (or ``off'') states on a timescale of hours that is atypical for what would be considered normal nulling behavior, which typically occurs on timescales of $\lesssim 100P$ \citep{2007MNRASWang}.
Analysis from both \citet{2014MNRASYoung} and \citet{2016MNRASHobbs} show that, in general, the duration of a typical bright state is 1--45 minutes, while the fraction of time during which PSR \psr\ is actually detected in the bright state versus the observing duration (i.e. its ``bright-state duty cycle'') is $\delta\sim 5\text{--}8\%$.
Compared to the other intermittent pulsars, which exhibit ``off'' times scales between $\sim 1\text{--}10^4$\,hr, PSR \psr\ is the second-most active intermittent pulsar (after PSR J1717--4054), but still poses a significant observational challenge.

In this paper, we present an analysis of simultaneous observations of PSR \psr\ with the Murchison Widefield Array (MWA; \citealt{2013PASATingay}) and the recently upgraded Molonglo Observatory Synthesis Telescope (UTMOST; \citealt{2017PASABailes}). 
The MWA is a low-frequency (70--300\,MHz) Square Kilometre Array precursor telescope located in Western Australia at the Murchison Radioastronomy Observatory. 
We used the high time resolution Voltage Capture System (VCS; \citealt{2015PASATremblay}) to record the low-frequency data.
UTMOST monitors many pulsars on a regular basis and has proven to be a capable detector of Fast Radio Bursts (e.g. \citealp{2017MNRASCaleb,2018MNRASFarah}).
We used UTMOST to record the higher-frequency (835\,MHz) data. 
In Section~\ref{sec:observations}, we describe the observations, post-processing and flux density calibration. We present our results and discuss them in Section~\ref{sec:results_and_discussion}. 
Low-frequency detection prospects are discussed in Section~\ref{sec:det_prospects}, and we present our summary and conclusions in Section~\ref{sec:conclusion}.
Throughout, we define the spectral index $\alpha$ as $S_\nu \propto \nu^\alpha$, $S_\nu$ as the flux density at frequency $\nu$, and $F_\nu=\int S_\nu\,{\rm d}t$ as the fluence at frequency $\nu$, which acts as a proxy to the pulse energy.

\section{Observations and Data Processing}\label{sec:observations}
In 2017 April, we began an observing campaign with the MWA and UTMOST to simultaneously observe PSR \psr. 
This amounted to 10 contemporaneous observations over a period of six months, during which not only did the MWA change configurations, but UTMOST also converted into a transit instrument.
These observations were facilitated by the real-time processing capabilities of UTMOST and the triggering capability (newly developed in the case of the MWA, see Section~\ref{sec:vcs_buffer}) of both instruments. 
For all but one observation, personnel at both instruments were required to monitor the telescopes for the duration of the observing runs (typically $\sim 30\text{--}40$ minutes), and in the case of the MWA, trigger data recording to mitigate storage concerns (see Section~\ref{sec:mwa_obs}).
The initial observing runs were before UTMOST was converted into a transiting telescope; thus on 2017 April 26 (MJD 57869), we observed PSR \psr\ for $\sim 5$ minutes every 30 minutes for a total of 8 times with UTMOST.
Monitoring was not possible on 2017 September 3 (MJD 57999), so a standard pulsar observation was scheduled on the MWA to coincide with when the pulsar would be transiting UTMOST.
The final observation of the campaign was conducted on 2017 December 2. 
An overview of the observations can be found in Table~\ref{tab:obstrials}.

\begin{deluxetable*}{lcCccc}
\tablecaption{Simultaneous observation attempts\label{tab:obstrials}}
\tablehead{\colhead{MJD} & \colhead{UTC} & \colhead{Duration} & \colhead{Bright state} & \colhead{Single pulse(s)} & \colhead{Coincident?} \\
 & & \colhead{(min)} & \colhead{detected} & \colhead{detected} &}
\startdata
57865\tablenotemark{a} & 2017 Apr 22 & 5 & Yes & -- & No\\
57869 & 2017 Apr 26 & 8\times 5 & No & No & --\\
57970 & 2017 Aug 5 & 30 & No & Yes (2) & Yes (1)\\
57977 & 2017 Aug 12 & 30 & No & No & --\\
57984 & 2017 Aug 19 & 30 & No & No & --\\
57995 & 2017 Aug 30 & 30 & No & Yes (1) & No\\
57997 & 2017 Sep 1 & 30 & No & Yes (2) & No\\
57998 & 2017 Sep 2 & 30 & No & No & --\\
57999 & 2017 Sep 3 & 30 & Yes & Yes & Yes\\
58089 & 2017 Dec 2 & 30 & No & No & --\\
\enddata
\tablecomments{The numbers in parenthesis in the last two columns indicate the number of events for that observation. The pulsar was detected in its bright state on MJD 57999, hence the number of single-pulse events are not recorded here.}
\tablenotetext{a}{The MWA was unable to point correctly for this attempt. UTMOST has a marginal bright state detection lasting $\sim 5$ minutes. Single-pulse data were not available for this observation, thus we cannot comment on whether single pulses were detected.}
\end{deluxetable*}

On 2017 September 3, the MWA VCS recorded $\approx 1.4$\,hr of data at a central frequency 154.24\,MHz with a bandwidth of 30.72\,MHz, and UTMOST recorded $\approx 0.5$\,hr of data at a central frequency 835.59\,MHz with 31.25\,MHz bandwidth as the target transited.
This observation is the primary focus of this work.
Observation details of the bright state detection are summarized in Table~\ref{tab:obsparams}.

\begin{deluxetable*}{lcc}
\tablecaption{Observation parameters on 2017 September 3\label{tab:obsparams}}
\tablehead{\colhead{Parameter} & \colhead{MWA} & \colhead{UTMOST}}
\startdata
Center frequency (MHz)     		& 154.24 & 835.59 \\
Bandwidth (MHz)            		& 30.72  & 31.25 \\
System temperature (K)			& 473 	 & 278\tablenotemark{a} \\
Gain (K Jy$^{-1}$)				& 0.24 	 & 2.3 \\
FWHM (arcmin)              		& $\approx 2.8$	& $0.77\times 168$ \\
Time resolution ($\mu$s)   		& 100 	 & 655.36 \\
Frequency resolution (MHz) 		& 0.01 	 & 0.78125 \\
UTC start time					& 02:05:09 & 02:05:56 \\
Observations duration (s)		& 5154 	   & 1799 \\
\hline
Dispersion smearing in lowest channel (ms)\tablenotemark{b} & 1.26    & 0.48 \\
Dispersion delay across bandwidth (ms)\tablenotemark{b}     & 2887.81 & 18.12 \\
Dispersion delay between observed bands (ms)\tablenotemark{b,c} & \multicolumn{2}{c}{8532.06} \\
\enddata
\tablenotetext{a}{See Section~\ref{sec:utmost_flux} for details regarding the flux density calibration of UTMOST.}
\tablenotetext{b}{Assuming a dispersion measure of $40.75\,\mathrm{pc\,cm^{-3}}$ -- see Section~\ref{sec:dm} for details.}
\tablenotetext{c}{Delay between the top of UTMOST band and the bottom of the MWA band.}
\end{deluxetable*}

\subsection{UTMOST}\label{sec:utmost_obs}
The Molonglo Observatory Synthesis Telescope \citep{1981PASAuMills,1994PASAuLarge} has recently been refurbished in the UTMOST project \citep{2017PASABailes}. 
It now operates in a 31.25\,MHz band at 835\,MHz, with a single circular polarization (PSR/IEEE right-hand circular; see \citet{2010PASAvanStraten}), with 512 narrow ($\approx 46$\,arcsec) fanbeams tiling a wide ($ 4.25\times 2.8$\,deg) field of view.
Since 2017 June/July, the telescope has been operating as a transiting instrument only, in order to reduce the stress placed on the mechanical phasing mechanism during the large number of daily pointings in the pulsar timing program, which previously made maintenance infeasible.
This strategy has proved very effective, with an approximate doubling of the sensitivity, and very significant improvements in the phase and sensitivity stability of the telescope over time.

PSR \psr\ was typically observed for approximately 30 minutes around the meridian transit of the source.
At its decl., the source transits the $\sim 4^\circ$ primary beamwidth in 31 minutes.
The UTMOST real-time pulse detection system, which runs \textsc{Heimdall}\footnote{\url{https://sourceforge.net/projects/heimdall-astro/}}, reports pulse candidates with a signal-to-noise ratio ${\rm S/N}>10$ while operating.
Additionally, UTMOST is able to produce coherently de-dispersed pulses from voltages and fold the pulsar time series in real time, allowing alerts to be transmitted within 60\,s regarding whether the source is active or if there have been single-pulse events.

The UTMOST backend writes SIGPROC\footnote{\url{http://sigproc.sourceforge.net/}} filterbank format files to disk with 327\,$\mu$s time resolution resolution for 320 frequency channels at a frequency resolution of 98\,kHz.
In normal operations, these are decimated to 655 $\mu$s and $40 \times 0.78$\,MHz channels before being archived, but the decimation process was interrupted for the majority of observations and the high-resolution data were recorded.
After the bright-state detection in September, the pre-decimation data were not retrieved, so we instead used the coarser-resolution data.
For all observations of PSR \psr\ (excluding 2017 April 22), the data were incoherently de-dispersed and subdivided into single-pulse data files (``archives'') using the \textsc{dspsr} pulsar data processing software \citep{2011PASAvanStraten}.

\subsubsection{Flux density calibration}\label{sec:utmost_flux}
UTMOST consists of 352 telescope elements (``modules''), which have performance variations with time. 
As part of regular operations, the performance of each module is tracked when phase and delay calibration is performed (typically daily), allowing estimates of the overall system performance from day to day. 
Additionally, the overall system equivalent flux density (SEFD) is calculated by using a set of 10 well-calibrated, nominally unpolarized pulsars (including PSR J1644--4559) with accurate flux densities at 843\,MHz, as determined by \citet{2018MNRASJankowski}.
The full set of calibrator pulsars is typically observed over the course of a few weeks (F. Jankowski et al. submitted; V. V. Krishnan et al. in prep.). 
These approaches are independent means for the long-term performance of the system to be monitored. 
At the time of the primary bright state detection of PSR \psr\ in this paper (2017 September), the SEFD of UTMOST is estimated to be $\sim 120$\,Jy with an uncertainty of $\sim 50\%$.

\subsection{MWA}\label{sec:mwa_obs}
The MWA was originally composed of 128 tiles distributed with a maximum baseline of $\sim 3$\,km \citep{2013PASATingay}, nominally referred to as MWA Phase I.
MWA Phase II, which began operations in early 2017, includes an additional 128 tiles: 76 in two redundant hexagonal (``Hex'') configurations near the array core, and the remaining 52 tiles spread out to provide maximum baselines of $\sim 6$\,km \citep{2018arXivWayth}. 
However, the MWA signal processing chain is currently only able to ingest data from 128 tiles at a time.
The Phase II MWA is therefore periodically reconfigured into either a \textit{compact} configuration, where the ``Hex'' tiles are connected with the core of the array, out to $\sim 300$\,m from the array center, or an \textit{extended} configuration, where the long baseline tiles and some fraction of the core (excluding the ``Hex'' tiles) are connected.
All data examined in this campaign were collected in Phase II of the MWA in both the compact and extended configurations.
Each tile consists of 16 evenly spaced dipole antennas in a regular 4\,m$\times$4\,m grid.
The MWA can record 30.72\,MHz instantaneous bandwidth and observes in the frequency range $70\text{--}300$\,MHz.

The VCS provides the high time and frequency resolution observing mode for the MWA, capable of capturing the tile voltages after the polyphase filterbank channelization stage within the standard MWA signal processing pipeline (see \citealp{2015PASATremblay}).
This allows us to record critically sampled complex voltage streams from each tile (at 100\,$\mu$s time resolution, 10\,kHz frequency resolution) from each of the 24$\times$1.28\,MHz ``coarse'' channels to on-site disks at a data rate of $\sim 28\,\mathrm{TB\,hr^{-1}}$.
For all of our observations, we observed with a contiguous 30.72\,MHz bandwidth (i.e. all 24 coarse channels are adjacent) at a center frequency of 154.24\,MHz.

\subsubsection{Buffered VCS recording}\label{sec:vcs_buffer}
One of the difficult aspects of observing this pulsar with the MWA is the inherent limitation to the amount of data we can record at any time (storage capacity is reached after $\approx 100$ minutes).
To mitigate this, we observed the pulsar with the VCS in a bespoke buffer mode (P. Hancock et al. in prep.), where rather than writing data to disk, the voltages are kept in memory for as long as possible.
When we receive a trigger from UTMOST that the pulsar was either in its bright state or emitting single pulses, we dump those voltages in memory onto the disks while continuing to record new data.
This effectively gives the VCS the ability to record voltages from approximately three minutes prior to the actual trigger time.
The buffered VCS mode is still under development, but we were able to use an early version of this recording mode between 2017 April--December, albeit with significant human interaction (triggers from UTMOST were sent manually and the VCS recording also needed to be started and stopped manually).
Using this prototype stage of buffered recording, we triggered voltage capture in three instances (out of 10 observations) when UTMOST detected single pulses from PSR \psr\ (see Table~\ref{tab:obstrials}).

For the observation on 2017 September 3, when the pulsar was in its bright state, we were not operating in this buffer mode and had scheduled a ``normal'' VCS observation (1.5\,hr) to begin recording $\sim$1 minute before the target entered the UTMOST beam.
Regardless of the VCS recording mode, the captured voltages are in the same format and thus need to be post-processed and calibrated using the standard pipeline.

\subsubsection{Tied-array beamforming}\label{sec:mwa_beamforming}
For the MWA-VCS, a tied-array (or coherent) beam is formed in post-processing by summing the individual tile voltages in phase and then detecting the power (see e.g. \citealp{2016ApJBhat,2017ApJMeyers}; S. Ord et al. submitted). 
This reduces the field of view to approximately the synthesized beam of the telescope ($\sim 1$\,arcmin in the extended configuration and $\sim 20$\,arcmin in the compact configuration at 200\,MHz), but grants a significant improvement in S/N over that provided by the incoherent sum (i.e. summing the tile powers directly).
While the incoherent sum theoretically affords a sensitivity boost of $\sqrt{N_{\rm tiles}}$, where $N_{\rm tiles}$ is the number of tiles use to create the sum, the coherent beam increases the theoretical sensitivity by approximately $\sqrt{N_{\rm tiles}}$ again.
 
A tied-array beam is created by an offline post-processing pipeline implemented at the Pawsey Supercomputing Centre\footnote{\url{https://www.pawsey.org.au/}} on the Galaxy cluster.
The pipeline involves combining individual tile responses, cable and geometric delays, and complex gain information for each tile, per frequency channel, based on calibration solutions from the Real Time System (RTS; \citealp{2008ISTSPMitchell}).
In the case of the bright-state detections, the calibration model was produced from an observation of Pictor A (PKS 0518--45), approximately 4\,hr before the start of the observation.
After the recent reconfiguration of the MWA, data from 38 of the 128 tiles were excised due to poor calibration solution quality and/or unreliable dipole elements.
For all observations, the weightings applied to each tile to form the tied-array beam are then determined by minimizing the $\chi^2$ error between the target data and a sky model based on the calibrator solutions.
The recorded voltages were de-dispersed and subdivided into single-pulse archives using \textsc{dspsr}.

\subsubsection{Flux density calibration}\label{sec:mwa_flux}
Flux density calibration was realized using the method described by \citet{2017ApJMeyers}, which we briefly summarize below.
The flux density estimation is carried out by simulating the tied-array beam pattern and combining it with the tile beam model \citep{2015RaScSutinjo,2017PASASokolowski}, thereby allowing the gain ($G$) and system temperature ($T_\mathrm{sys}$) to be calculated.
This involves integrating over the tied-array beam pattern (to determine the gain) and then over the product of the tied-array and tile beam patterns with a given sky temperature map (to determine the sky temperature). 
We elected to use the Global Sky Model of \citet{2008MNRAS_GSM}, evaluated at 154.24\,MHz, as our sky temperature map.
The simulations assume that the tied-array beamforming process is ideal and that we should see an increase in S/N by a factor of $\sqrt{N_\mathrm{tiles}}$, as previously explained.
However, in general, the theoretical improvement is not achieved due to a combination of factors including calibration quality, the beamforming process, and the beam models employed.
At the time of the bright-state detection with UTMOST, we achieved $\approx 60\%$ of the theoretical improvement.
Using the estimated gain and system temperature, and taking into account the above considerations (see equation 2 of \citealp{2017ApJMeyers}), the SEFD during the bright state observation was $\sim 3300$\,Jy.

\section{Results and Discussion}\label{sec:results_and_discussion}
Throughout this Section, we focus on the simultaneous bright-state detection on MJD 57999 (2017 September 3), unless otherwise noted in the text.

\subsection{Pulse finding and cross-matching}\label{sec:sp_search}
We used the \textsc{psrchive} \citep{2004PASAHotan,2012AR&TvanStraten} software suite to process the single-pulse archives produced by \textsc{dspsr}.
Radio-frequency interference (RFI) was mitigated using the \textsc{paz} routine by utilizing the built-in median-smoothed difference algorithm. 
For the MWA, we also excised the edge channels of each of the 24 coarse channels to mitigate the effects of aliasing introduced by the channelization process.
We then summed the archives for every pulsar rotation in polarization and frequency to create a time series of each rotation, ensuring they were re-binned to the same time resolution (256 bins, or $\Delta t\approx 987\mathrm{\,\mu s}$). 
To find individual pulses in these time series, we used the \textsc{psrspa} routine with a detection threshold of $6\sigma$ on the overlapping time when the pulsar was active for both telescopes, which corresponds to $\sim 1344$\,s.
The signal-to-noise ratio thresholds were then converted into flux density limits using the radiometer equation,
\begin{equation}
\sigma_\nu = \frac{\rm SEFD}{\sqrt{n_{\rm p}t_{\rm int}\Delta\nu}},
\end{equation}
where $\sigma_\nu$ is the $1\sigma$ noise measured in Jy at frequency $\nu$, $n_{\rm p}$ is the number of polarizations sampled, $t_{\rm int}$ is the integration time (in this case, because we are applying this to the measured S/N of each time sample in the time series, $t_{\rm int}=\Delta t\approx 987\mathrm{\,\mu s}$), and $\Delta\nu$ is the bandwidth.

A special consideration needs to be made for UTMOST, given that it only samples a single polarization ($n_{\rm p}=1$).
For polarized sources, depending on the degree of polarization, this means that the measured flux densities can be significantly inflated or reduced.
PSR \psr\ is a moderately polarized pulsar, and from archival Parkes data at 3.1\,GHz (a subset of that used by \citealp{2014MNRASYoung}) the bright-state emission circular polarization fraction is $V/I\approx -0.1$.
This means that UTMOST flux densities need to be scaled by a factor of 0.9 for the analysis herein (see Appendix~\ref{sec:appendix} for details on computing the scaling factor).
Additionally, due to the bandpass shape and RFI excision at UTMOST, the estimated S/N for pulses is underestimated by $\sim 25\%$, thus we apply another scaling factor of 1.25 to correct for this when converting to flux density units.
Bandwidth considerations need to also be made for the MWA data, given that 10 of the 128 fine channels (each 10\,kHz wide) are zero-weighted on each side of all coarse channels to avoid aliasing effects.
This culminates in reducing our effective bandwidth to $\sim 70\%$ of the full 30.72\,MHz, ergo the MWA flux densities are scaled by a factor of $(0.7)^{-1/2}\approx 1.2$ to correct for this.

Taking into account the above considerations, the nominal flux density limits are then $6\sigma_{154}\approx 97$\,Jy and $6\sigma_{835}\approx 4.6$\,Jy for the MWA and UTMOST, respectively.

After the automated pulse finding (using data that had been processed with an updated ephemeris; see Section~\ref{sec:dm}), each candidate pulse was visually inspected to ensure the validity of the detection, resulting in the identification and subsequent removal of 1 MWA and 17 UTMOST spurious events that were narrowband and likely RFI.
In total, \mwaN\ pulses were detected at the MWA, and \utmostN\ with UTMOST above the $6\sigma$ level\footnote{We note that, for UTMOST, there are 20/\utmostN\ pulses with peak flux densities between 3.3--4.3\,Jy, which is slightly less than the nominal limit (i.e. $\geq 4.3\sigma$).}.
We created a catalog of these detections for each telescope and cross-matched them based on their assigned ``pulse number'' using the STILTS software package \citep{2006ASPCTaylor}. 
The pulse number is computed as the number of rotations since some arbitrary time (for us, MJD 53089.00000), which is defined in the pulsar ephemeris.
Thus, the same rotation of the pulsar can be compared at each telescope. 
After accounting for the dispersive delay between the telescopes, \simN\ pulses were simultaneously detected with the MWA and UTMOST, i.e. a match rate (based on the MWA population) of $60\%$.

The same processing steps were applied to the other three simultaneous observations (MJD 57970, 57995, and 57997 in Table~\ref{tab:obstrials}) when UTMOST detected single pulses from PSR \psr\ while it was in its weak state.
In total, five single pulses were detected over those three separate observations. 
For one of the two single pulses detected at UTMOST on MJD 57970 (2017 August 5), there was a marginal simultaneous single-pulse detection from the MWA.
This single-pulse was detected in the UTMOST observation with ${\rm S/N}\sim 14$, and was consequently detected (with a ${\rm S/N}\lesssim 5$) by eye when looking at the corresponding pulsar rotation number in the MWA data.
There were no other coincident detections of weak state single pulses with the MWA.

\subsection{Dispersion measure}\label{sec:dm}
Propagation effects imposed on pulsar signals by the interstellar medium (ISM) are much stronger at low radio frequencies (e.g. dispersive delays scale as $\nu^{-2}$). 
To that end, measurements of the dispersion measure (DM) of a pulsar from even a single observation at low frequency can often be equally, if not more, precise compared to that obtainable using months of timing data at higher frequencies from larger telescopes. 
These refinements are important for understanding the ISM along the line of sight to the pulsar and for pulsar timing experiments at higher frequencies (e.g. \citealp{2014ApJPennucci,2015ApJLam,2017MNRASLentati}), where any errors in the DM, due to the frequency lever-arm or time variability, are not obvious from single observations.

\begin{figure*}
\centering
\includegraphics[width=\textwidth]{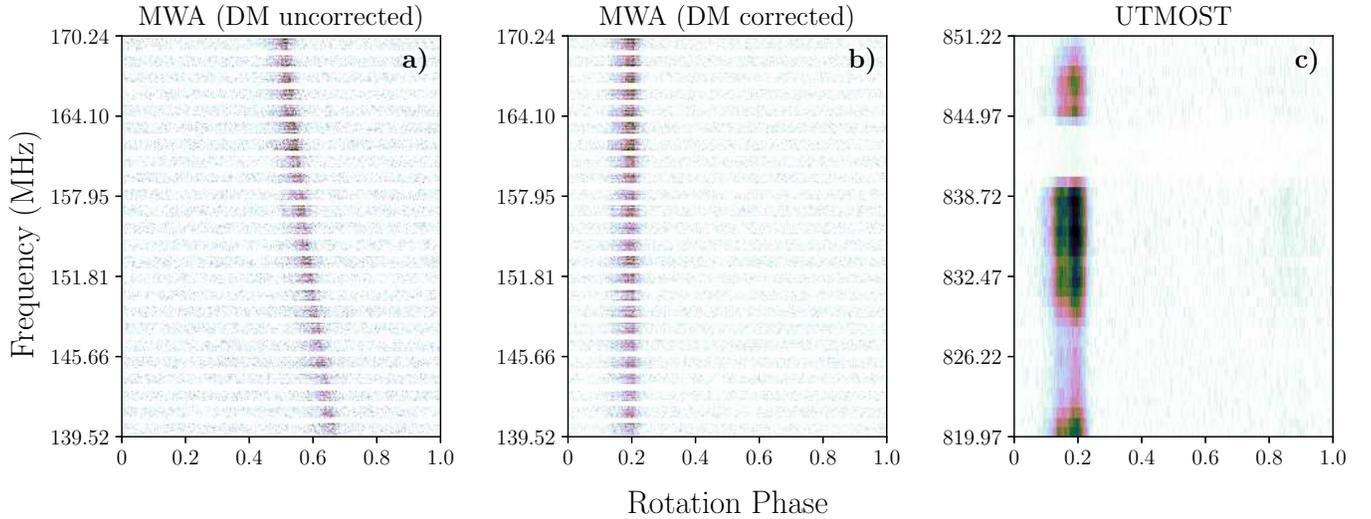}
\caption{Frequency vs. phase waterfall plot of the signal arriving at both telescopes. 
The panels are as follows: (a) the addition of all pulses detected with a signal-to-noise above 6, de-dispersed to the nominal cataloged dispersion measure ($40.2\mathrm{\,pc\,cm^{-3}}$); (b) the MWA data again, de-dispersed with the optimal DM output by a \textsc{pdmp} search ($40.75\mathrm{\,pc\,cm^{-3}}$), and; (c) the UTMOST data de-dispersed with a DM of $40.75\mathrm{\,pc\,cm^{-3}}$.
The horizontal stripes in the MWA images are where edge channels have been flagged for each of the 24 coarse channels. 
For the UTMOST plot, a fraction of the bandwidth around 841\,MHz was removed due to RFI caused by mobile phone networks, and the reduced signal around 826\,MHz is due to the instrument bandpass.
\label{fig:dmsmear}}
\end{figure*}

We averaged together single pulses above a detection significance of $6\sigma$, using the cataloged dispersion measure ($\DMcat \mathrm{\,pc\,cm^{-3}}$; \citealp{2014MNRASYoung}).
There was significant residual frequency-dependent delay of the arrival of pulses in the MWA data, corresponding to an excess time delay across the observed 30.72\,MHz bandwidth of $\approx 39$\,ms, which is $\sim 15\%$ of the pulse period (see Figure~\ref{fig:dmsmear}a).
The delay in the MWA band is dispersive, exhibiting the classical quadratic sweep, which is expected for an offset from the true DM.

To determine a more precise DM, we made use of the \textsc{psrchive} routine \textsc{pdmp} on the MWA data, which calculated the optimal DM to be $\DMobs \mathrm{\,pc\,cm^{-3}}$.
While this is within the uncertainty associated with the original and cataloged value, ${\rm DM_{cat}}=\DMcat \mathrm{\,pc\,cm^{-3}}$, a correction of $\delta {\rm DM}=\DMoffset \mathrm{\,pc\,cm^{-3}}$ is indeed significant, as is the factor of $\sim 50$ improvement in precision.
This DM offset introduces a delay across the UTMOST bandwidth of $\sim 1/4$ of a time sample ($\sim 0.25$\,ms), and hence it was not discernible from the UTMOST data (see Figure~\ref{fig:dmsmear}).
The revised DM results in a delay between the top of the UTMOST band and bottom of the MWA band of approximately $8532$\,ms (see Table~\ref{tab:obsparams}).

We note that, while this level of precision in DM is impressive from a single observation, we are only using the brightest pulses from the $\sim 20$ minutes of data containing the bright state and do not account for any potential profile evolution, nor the possibility of a frequency-dependent (chromatic) DM \citep{2016ApJCordes,2017MNRASShannon}.
Nonetheless, for our purposes, the updated DM produces higher signal-to-noise ratios for both the profiles and single pulses, and aligns the profiles in phase without any other alteration to the ephemeris used.

\subsection{Pulse profile}\label{sec:prof}
\begin{figure}
\centering
\includegraphics[width=0.5\textwidth]{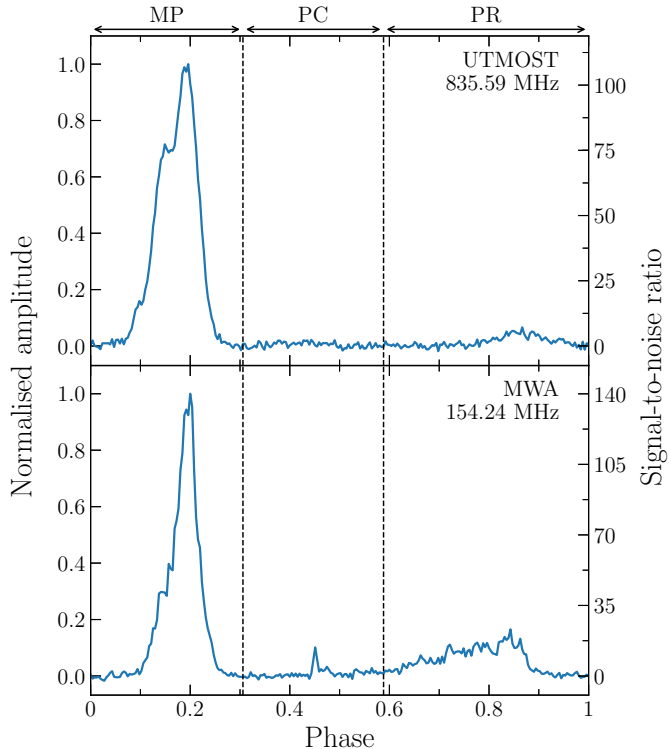}
\caption{A pseudo-integrated profile created by averaging all pulses with ${\rm S/N}>6$ of \psr\ from each telescope from the 2017 September 3 detections (\mwaN\ from the MWA, \utmostN\ from UTMOST).
The profile is split into three regions, corresponding to the components within, which have been labelled based on the classifications made by \citet{2014MNRASYoung}.
It appears that the postcursor (PC) and precursor (PR) components are more prominent at lower frequencies.
The PR emission is also shifted and has a significant rise time at low frequencies. \label{fig:profile_6sig}}
\end{figure}

After determining the DM offset, we applied the correction and re-processed the bright-state data, again conducting a single-pulse search as described in Section~\ref{sec:sp_search}.
Combining only those pulses with a detection significance $\geq 6\sigma$, as defined by the \textsc{psrspa} single-pulse finding algorithm from \textsc{psrchive} (\mwaN\ pulses for the MWA, \utmostN\ pulses for UTMOST), we form a ``pseudo-integrated'' profile\footnote{We note that this includes those UTMOST pulses that are nominally less than the $6\sigma$ flux density limit, which is ultimately due to the different ways in which the noise was estimated during single-pulse detections (automated off-pulse estimation) versus flux calibration (sigma-clipping).} (see Figure~\ref{fig:profile_6sig}).
This highlights the emission across a large portion of pulse longitude, hence Figure~\ref{fig:profile_6sig} is split into three phase regions, where we label these regions (left to right) with their corresponding profile component from \citet{2014MNRASYoung}: the ``main pulse'' (MP), ``postcursor'' (PC), and ``precursor'' (PR). 

\begin{figure}
\centering
\includegraphics[width=0.5\textwidth]{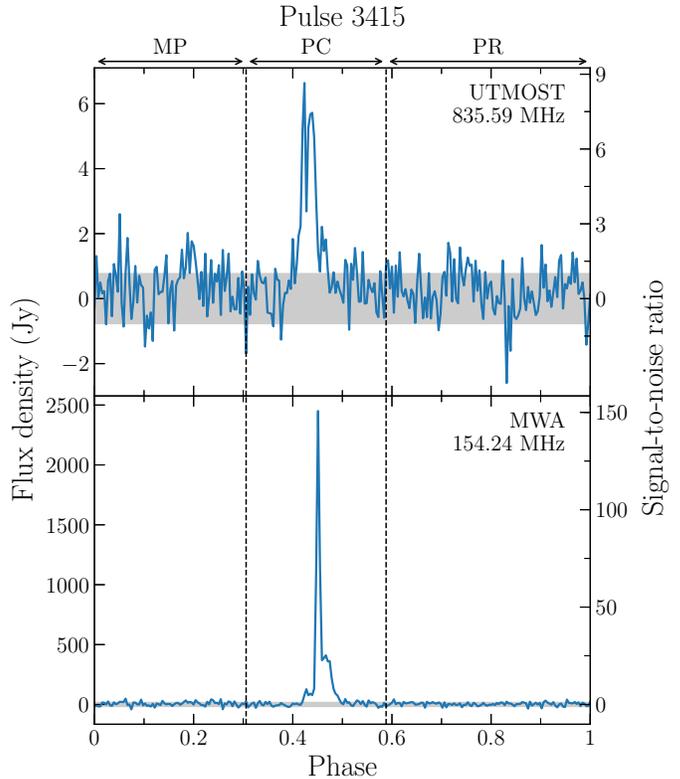}
\caption{A relatively weak PC-component pulse detected by UTMOST (top) with a peak flux density of $\approx 7$\,Jy, which is exceptionally bright at MWA frequencies (bottom), with a peak flux density of $\approx 2.5$\,kJy.
The grey-shaded region indicates the $\pm1\sigma$ noise level.
This pulse occurred 3415 rotations after the first simultaneously observed pulsar rotation. \label{fig:brightPC}}
\end{figure}

Even with the caveat that only a limited number of pulses contributed to these profiles, it is clear that there are some differences between the two frequencies.
In particular, the PR component appears brighter, wider, and shifted at 154\,MHz with respect to the 835\,MHz profile. 
At MWA frequencies, the PR component appears as a gradual rise from a phase of $\sim 0.6$ with a sudden cutoff around 0.9, whereas the equivalent UTMOST component only appears between $\sim 0.8\text{--}1$ in phase and is smoother overall.
Single pulses in the PC phase regions are recorded at both telescopes, but there was one exceptionally bright pulse in the MWA data (${\rm S/N}\sim 150$) that dominates in this case, whereas the PC single pulses from UTMOST were typically quite weak (see Figure~\ref{fig:brightPC}).
This bright pulse acts as an alignment anchor, indicating that the profile alignment is real and thus the differences in the profile features must be due to the emission mechanism(s) and/or magnetospheric propagation effects at play.

\subsection{Spectral index}\label{sec:spectral_index}
\begin{figure}
\centering
\includegraphics[width=0.45\textwidth]{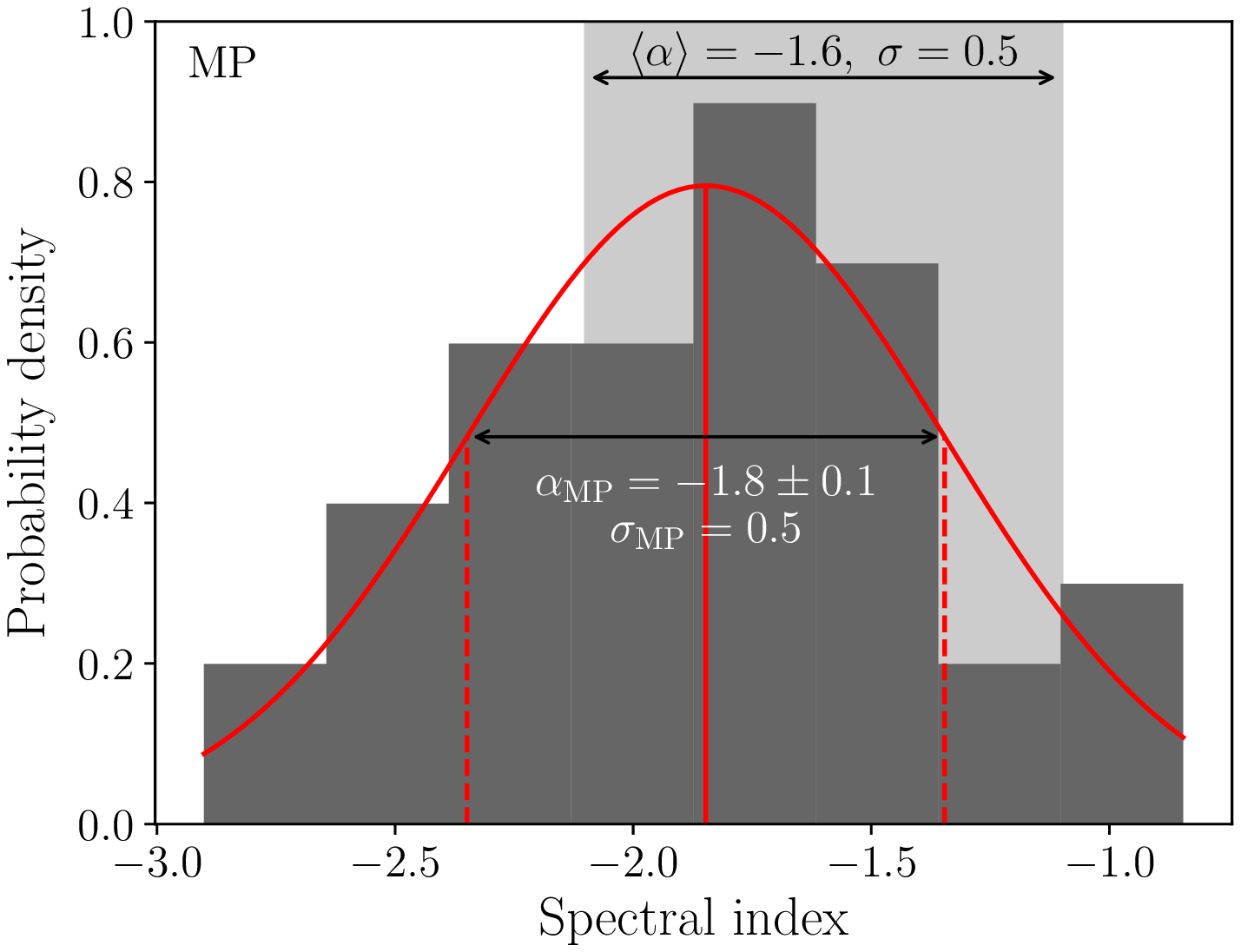}
\includegraphics[width=0.45\textwidth]{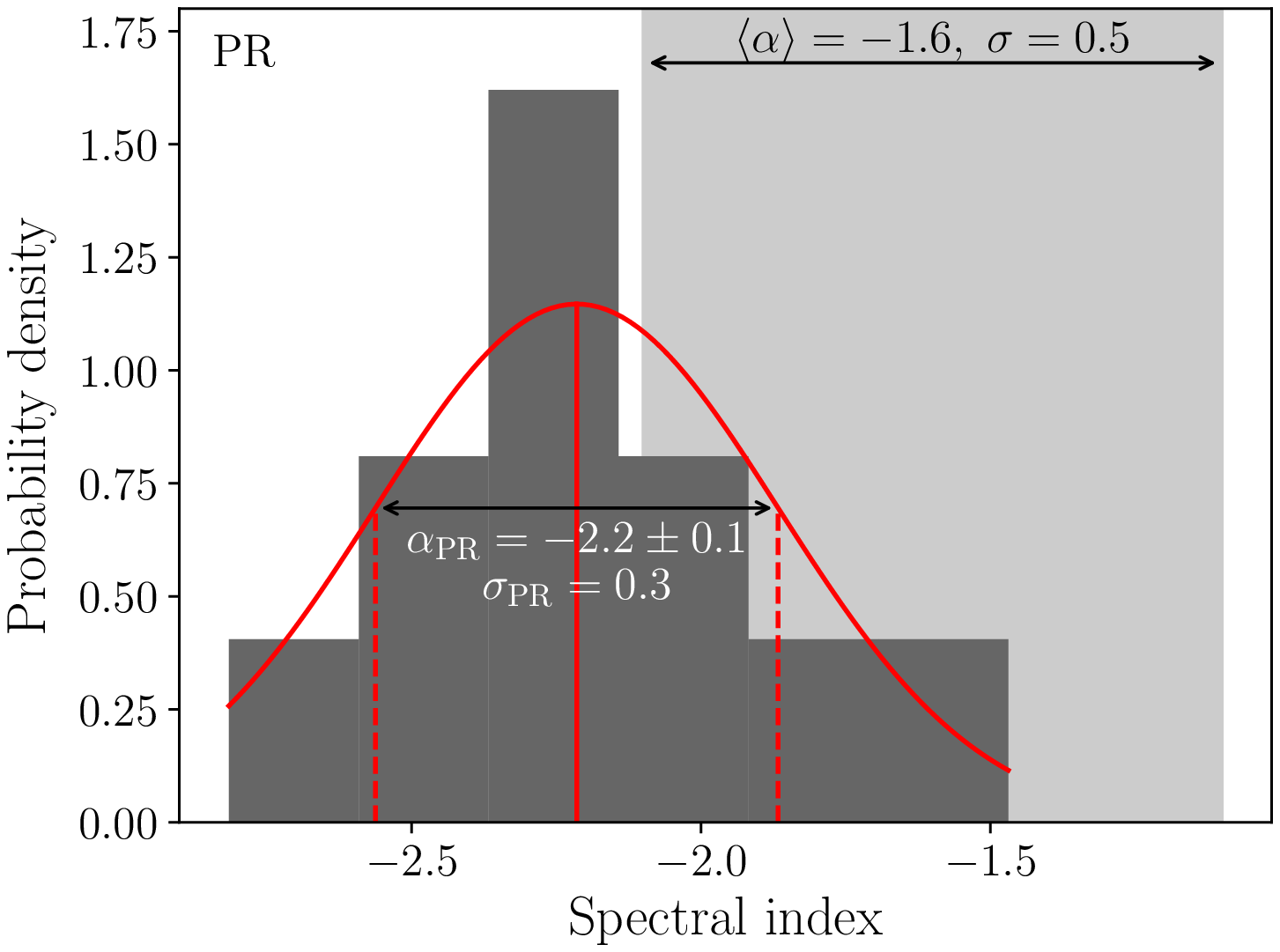}
\caption{Spectral indices between the MWA (154.24\,MHz) and UTMOST (835.59\,MHz) for cross-matched pulses, separated by component. 
The typical spectral index range (i.e. a mean of $\alpha=-1.6$ and standard deviation of $\sigma=0.5$) for pulsars, as determined by \citet{2018MNRASJankowski}, is shaded in the background to provide context.
The Gaussian fits are only indicative, given the small number of pulses contributing to each of the distributions (39 and 11 for the MP and PR components, respectively), but identify the mean spectral indices to be $\alpha_{\rm MP}=\alphaMP$ and $\alpha_{\rm PR}=\alphaPR$ for the MP and PR components, respectively.
\label{fig:alphas}}
\end{figure}

From the individually matched pulses (\simN\ in total), we calculated the spectral index distributions for the MP and PR components using their measured fluences.
For the MP component, the mean spectral index is relatively typical of the normal pulsar population, with $\alpha_{\rm MP}=\alphaMP$ and a standard deviation of \alphaMPstd.
The mean spectral index of the PR component is steeper than typical with $\alpha_{\rm PR}=\alphaPR$ and has a narrower distribution with a standard deviation of \alphaPRstd\ (see Figure~\ref{fig:alphas}).
The only pulse in the PC phase region with a counterpart (see Figure~\ref{fig:brightPC}) has an extreme spectral index of $\alpha_{\rm PC}\approx \alphaPC$.
We note that, given our definition of fluence, we are often summing over multiple components within the specified MP and/or PR windows (e.g. pulse 3773 in Figure~\ref{fig:spgallery}), which can act to bias the measured spectral index.

The spectral index distribution for the MP component is generally within the normal pulsar population spectral index distribution, which has a mean of $\langle\alpha\rangle=-1.6\pm 0.03$ and a standard deviation of $\sigma=0.5$ (\citealp{2018MNRASJankowski}; see also \citealp{2000A&ASMaron,2013MNRASBates}), as indicated by the shaded regions in Figure~\ref{fig:alphas}.
The steepness of the PR component spectral index could explain the disparity between the number of pulses detected in each phase range (see Section~\ref{sec:sp}), where the PR component will become brighter at lower frequencies faster than the MP component, thus somewhat equalizing our MP-to-PR ratio measured with the MWA.
However, scaling the detected MWA pulses in the PR phase range with the measured spectral index to the expected flux density at 835.59\,MHz, we find that only eight of the 43 pulses fall below the nominal UTMOST flux density limit.
Thus, there must be some other contributing factor as to why some of these pulses are detected at the lower, but not the higher, frequencies.

Based on a radius-to-frequency mapping argument, it is possible that, at low frequencies, the magnetospheric region producing the PR emission is actually distinct from, and possibly more active than, the higher frequency emission region.
The separate emission regions would therefore also experience different propagation paths through the magnetosphere along the line of sight, which could also act to suppress the higher-frequency emission in this region.

\subsection{Pulse energy distributions}\label{sec:fluence_dist}
Characterizing the pulse energy or amplitude distribution of a pulsar is useful in understanding the pulse emission process.
The pulse energy distributions of individual pulsars vary substantially, but are typically seen to be a log-normal (LN) or exponential distribution (e.g. \citealp{2012MNRASBurke-Spolaor}), or a power law (PL) in the case of giant pulses (e.g. \citealp{2008ApJBhat,2012ApJMickaliger,2017ApJMeyers}).
Few studies of this kind have been done for intermittent pulsars in general (e.g. \citealp{2015MNRASSobey}).
Here, we attempt to characterize the pulse energy distributions of the MP and PR components (see Figure~\ref{fig:profile_6sig}) from the MWA and UTMOST.

The pulse energy distribution of PSR \psr\ is hard to assess, given its intermittent nature, but there are a handful of examples where estimates have been made.
\citet{2012MNRASBurke-Spolaor} attempted to estimate the Gaussian and LN distribution parameters for a weak state detection of one pulse, so the results are not statistically significant.
\citet{2014MNRASYoung}, who have the largest sample of pulses from PSR \psr\ in all emission states in the literature, report that a PL distribution is the most appropriate fit during the bright state , while the weak state is better parametrized by a LN distribution.
However, the authors use the term ``pulse energy'' interchangeably with pulse intensity (i.e. flux density or amplitude), thus it is unclear whether we can directly compare distribution parameters. 
In any case, a LN model was not fit to the data, so we cannot compare those distribution parameters.
\citet{2018arXivMickaliger}, through reprocessing of PMPS archival data, also re-detect PSR \psr\ in a single-pulse search in the weak state.
The data did not provide sufficient statistical power to discriminate between the trial distributions (PL, LN and exponentially truncated power law (TPL)), which is to be expected given the small number of detections (18 out of 8300 rotations).

\begin{figure*}
\centering
\includegraphics[width=0.48\linewidth]{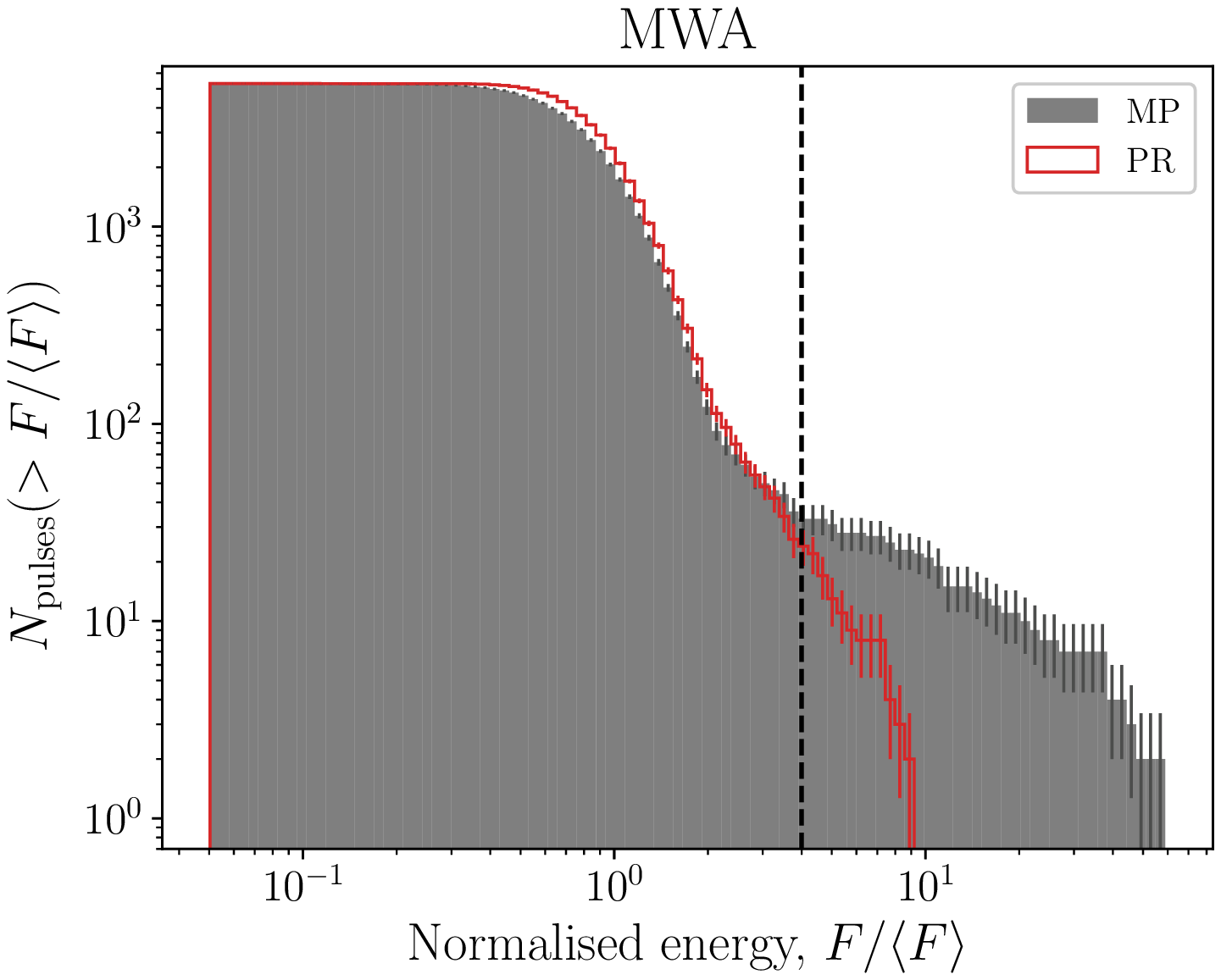}
\includegraphics[width=0.48\linewidth]{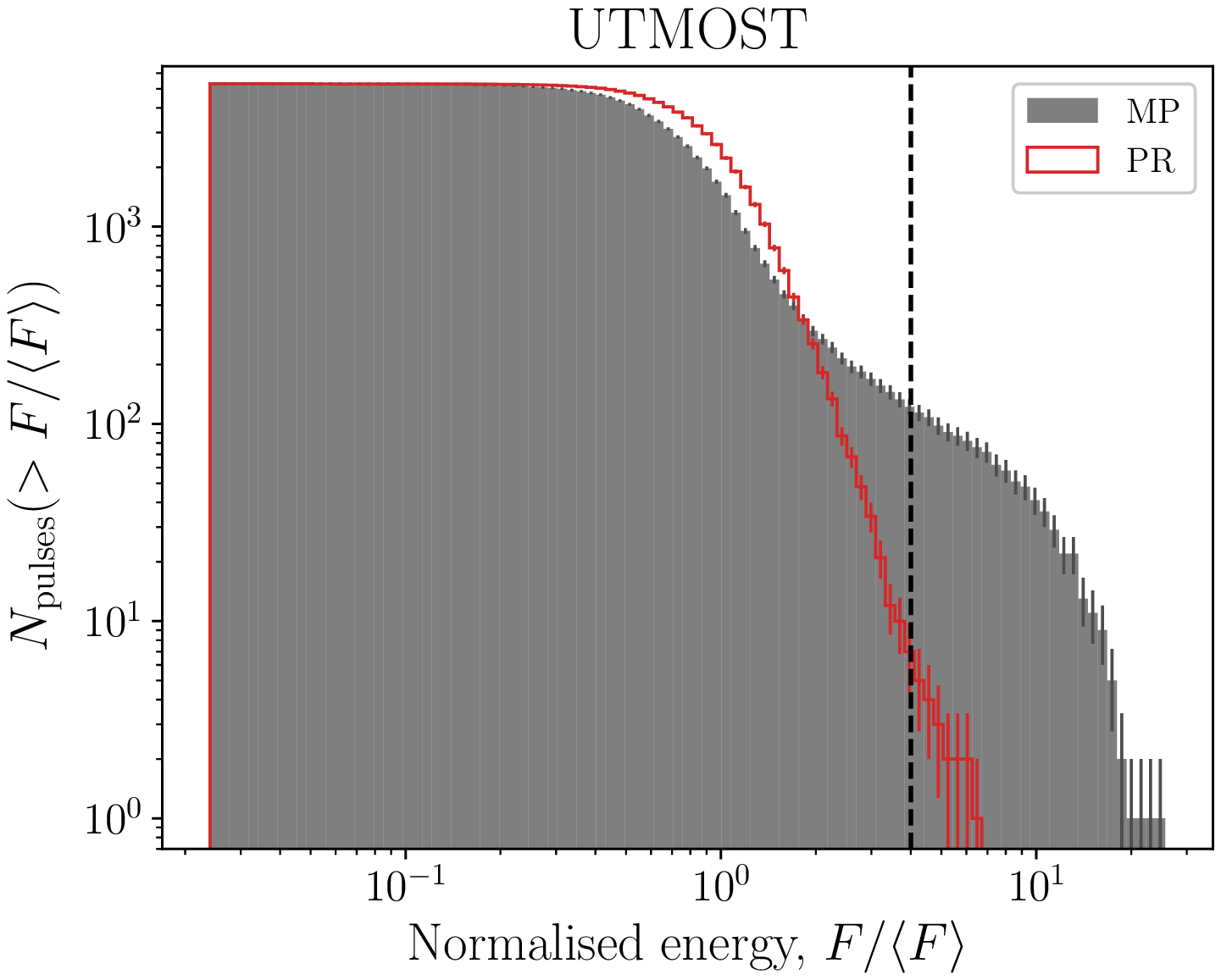}
\includegraphics[width=0.48\linewidth]{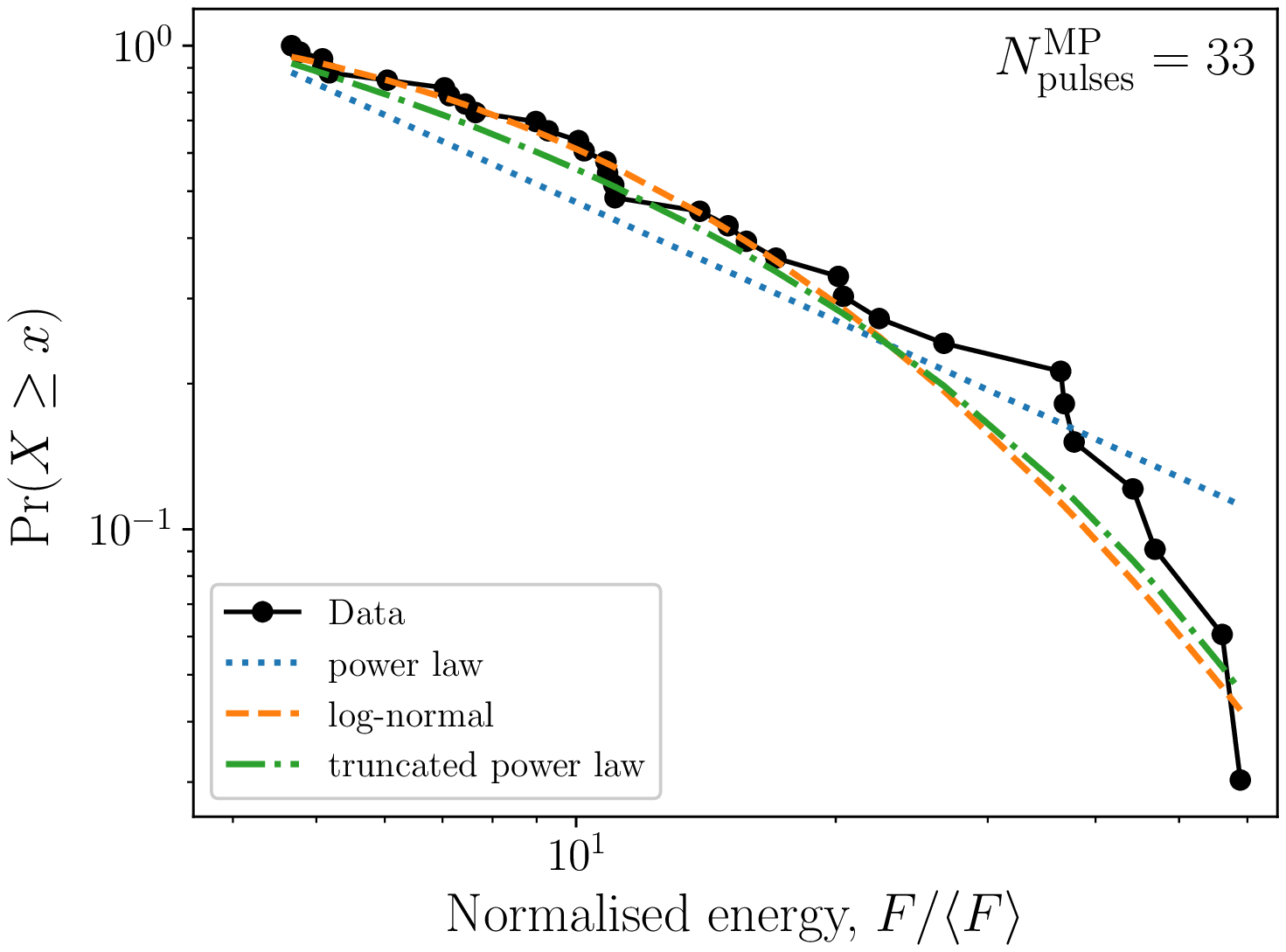}
\includegraphics[width=0.48\linewidth]{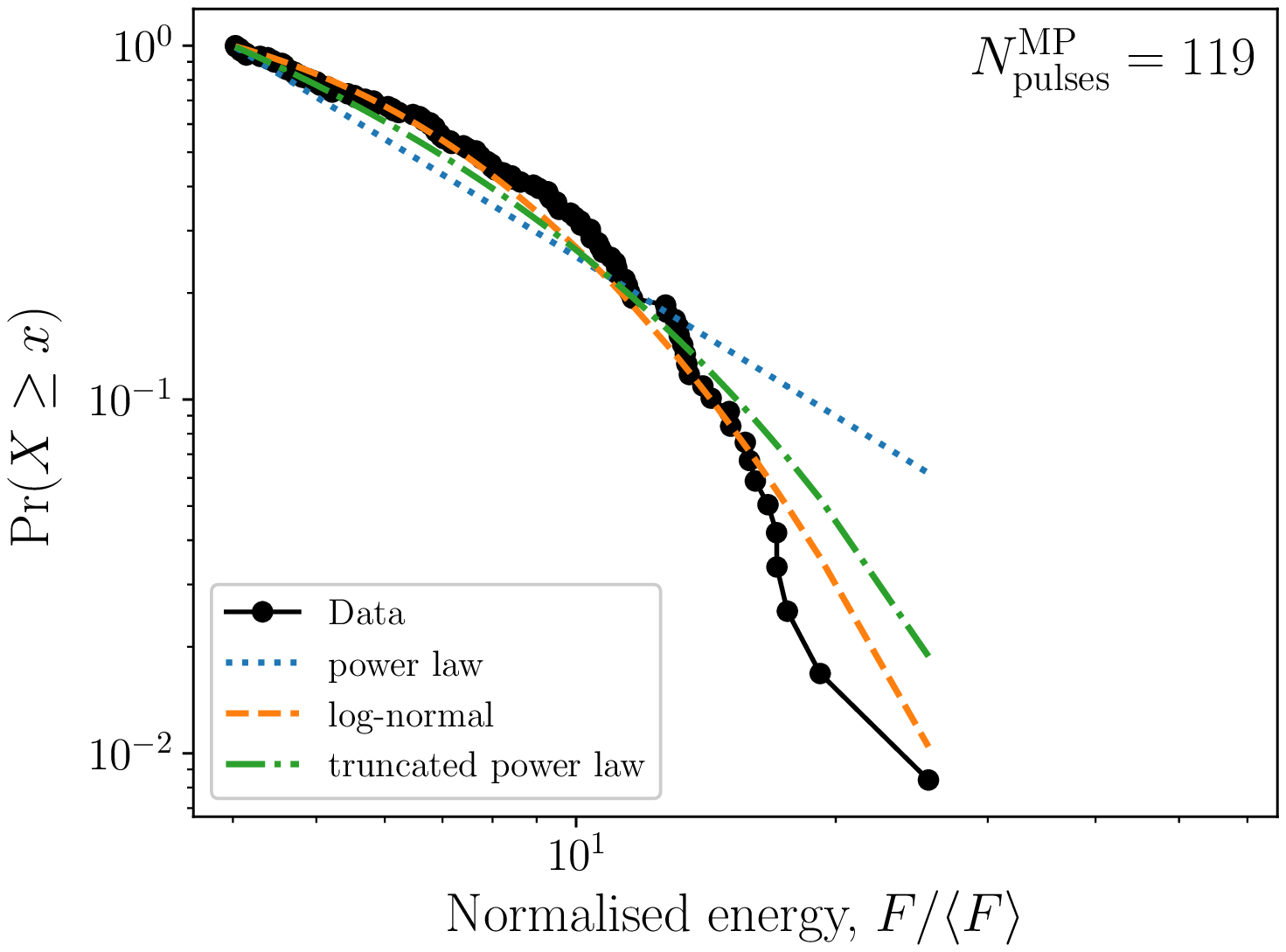}
\includegraphics[width=0.48\linewidth]{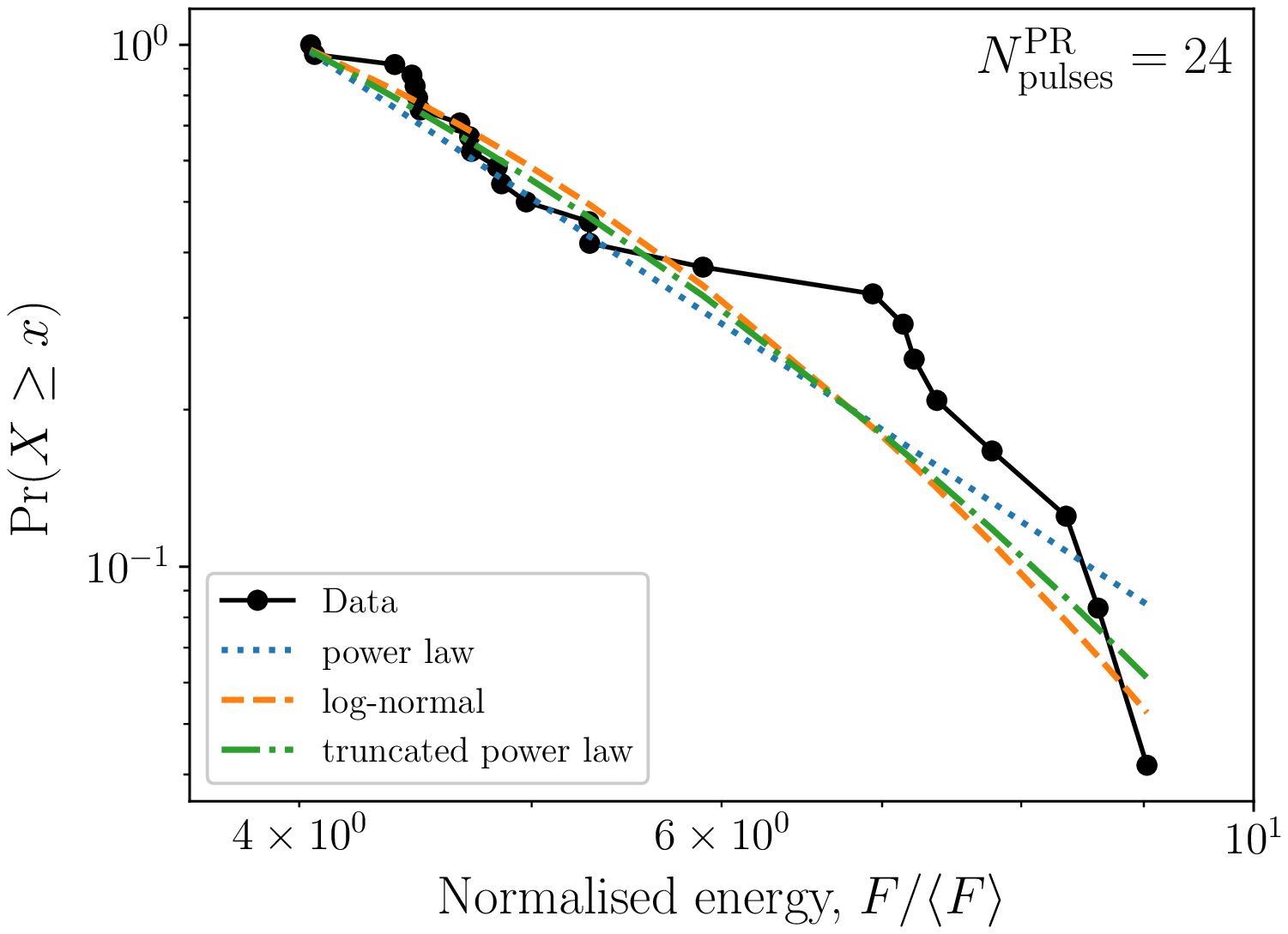}
\includegraphics[width=0.48\linewidth]{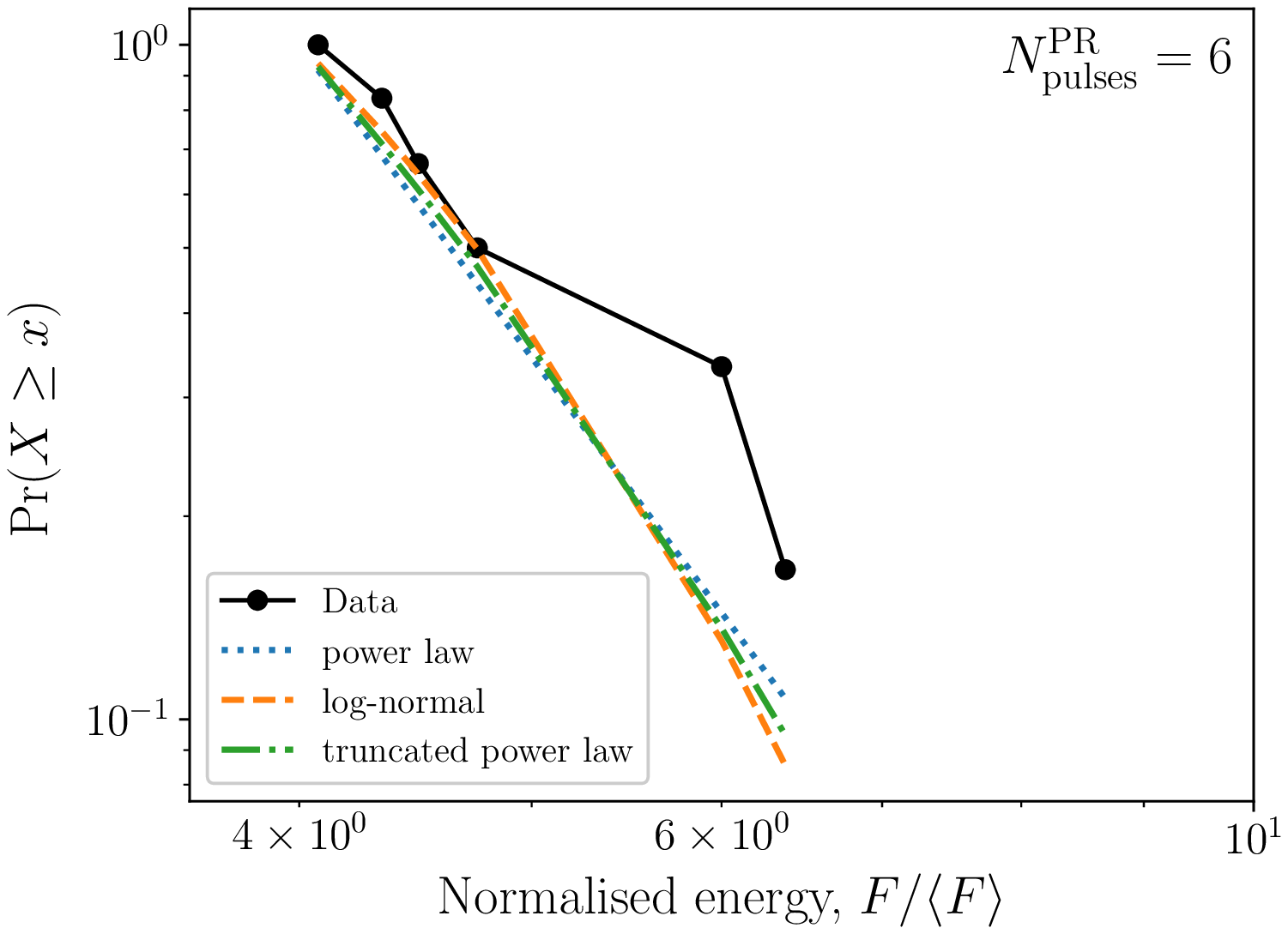}
\caption{Normalized fluence cumulative distributions for the MP and PR components from the MWA (left) and UTMOST (right).
The top row shows the number of pulses above a given normalized energy.
The middle and bottom rows show the distribution for those pulses in the MP and PR phase regions, respectively, with normalized fluences $F\geq 4\langle F\rangle$ (indicated by the vertical dashed line in the top row).
The best-fit models are also drawn (see Table~\ref{tab:bestfit-params}), though the PR component from UTMOST is not well constrained given only six data points.
For the middle and bottom rows, the ordinate is scaled to represent the fraction of pulses above a given normalized energy, i.e. the survival function, $\mathrm{Pr}(X\geq x)$.\label{fig:fluhist}}
\end{figure*}

We define the pulse energy (or ``fluence'') as the integrated flux density over the emission component window above the baseline fluctuations.
During the integration, the baseline noise was estimated independently for each pulsar rotation using the sigma-clipping method.
In our case, we integrate over each of the three phase regions defined previously for every pulsar rotation, regardless of whether there was a pulse detected, and take that as an estimate of the component fluence for that rotation.
We normalize our pulse energies by dividing each measurement by the average energy, $\langle F\rangle$, for each of the pulse components over the $\sim 5300$ rotations (see top row of Figure~\ref{fig:fluhist}).
The Python \textsc{powerlaw}\footnote{\url{https://github.com/jeffalstott/powerlaw}} module \citep{2014PLoSOAlstott} was used to fit distributions typically tested in the literature: a PL, LN, and TPL.
We also limited ourselves to only fit pulses with a normalized fluence $F\geq 4\,\langle F\rangle$, which we note is a somewhat arbitrary choice. 
In general, the \textsc{powerlaw} package can compute the appropriate cutoff by minimizing the Kolmogorov-Smirnov distance, but in this case, because we are so heavily dominated by the ``noise,'' the automatic estimation fails---so we elected to ensure that only real pulses are being included by setting a relatively conservative limit.
The respective non-normalized probability density forms of these distributions are given in equations~\eqref{eq:powerlaw}, ~\eqref{eq:lognormal}, and \eqref{eq:truncpowerlaw} below:
\begin{align}
P_{\rm PL}(x) & \propto x^{-\beta} \label{eq:powerlaw}\\
P_{\rm LN}(x) & \propto \frac{1}{\sigma x}\exp\left[-\frac{\left(\ln x-\mu\right)^2}{2\sigma^2}\right] \label{eq:lognormal}\\
P_{\rm TPL}(x) & \propto x^{-\Gamma}\exp\left(-\lambda x\right) \label{eq:truncpowerlaw}
\end{align}

To determine which of the distributions best parametrizes our data, we compare the log-likelihood ratios ($R$), initially with respect to a PL, and the corresponding significance ($p$-values), which are calculated as part of the fitting procedure.
In this case, negative values of $R$ favor the opposing model (i.e. not the PL).
Parameters for the best-fitting distributions of each kind are given in Table~\ref{tab:bestfit-params}, along with their associated $R$ and $p$-values with respect to a PL.
The best fits are shown in the middle and bottom rows of Figure~\ref{fig:fluhist}.

For the MP component, a LN distribution is favored for the UTMOST data ($R=-13.3$), while either a LN or TPL are statistically plausible for the MWA data.
To better determine which distribution is favored for the MWA overall, we compared the $R$ and $p$-values of a TPL with respect to a LN distribution, which indicated that the LN distribution may be a better fit ($R=0.1$), but the significance ($p=0.9$) is inconclusive.

We conducted the same fitting procedures for the PR component data and found that we cannot significantly discern which trialled model provides the best fit for either telescopes .
The fits to the MWA data slightly favor a LN or TPL distribution based on their $R$ values, however, we cannot statistically reject the PL based on the respective $p$-values.
As found for the MP component, a LN model is favored over a TPL ($R=0.02$), however the $p$-value here is also inconclusive ($p=0.6$).
For the UTMOST data, given there are only six data points, caution must be taken when interpreting the results, but the log-likelihood ratios tend to favor a LN distribution in all cases, including when compared to a TPL.

From a global perspective, it appears that a simple PL distribution is not appropriate for our data, with the caveat that we are limited by small number statistics.
Assuming that the pulse energy distribution does not change ``type'' as a function of frequency or component, we tend to favor a LN distribution for the MP and PR pulse energies.
In this respect, PSR \psr\ is like most normal pulsars (e.g. \citealp{2012MNRASBurke-Spolaor}).

\begin{deluxetable*}{lCCCCCCCC}
\tablecaption{Best-fitting parameters for the normalized fluence distributions.\label{tab:bestfit-params}}
\tablehead{\colhead{Component} & N_{\rm pulses} & \colhead{Power law (PL)} & \multicolumn{3}{c}{Log-normal (LN)} & \multicolumn{3}{c}{Truncated power law (TPL)}\\
 & (>4\langle F\rangle) &\colhead{$\beta$} & \colhead{$\mu$} & \colhead{$\sigma$} & \colhead{$(R/p)\tablenotemark{a}$} & \colhead{$\Gamma$} & \colhead{$\lambda$} & $(R/p)\tablenotemark{a}$}
\startdata
\sidehead{Main pulse (MP)}
MWA    & 33  & 1.81\pm 0.14 & 2.3\pm 0.5 & 1.0\pm 0.2 & -4.8/0.03   & 1.0\pm 0.1     & 0.03\pm 0.01 & -4.7/0.002\\
UTMOST & 119 & 2.49\pm 0.13 & 1.8\pm 0.1 & 0.6\pm 0.1 & -13.3/0.002 & 1.0\pm 10^{-5} & 0.12\pm 0.01 & -11.9/10^{-6}\\
\sidehead{Precursor (PR)}
MWA    & 24  & 4.0\pm 0.6   & 1.3\pm 0.4 & 0.4\pm 0.3 & -0.98/0.3   & 1.0\pm 0.7     & 0.4\pm 0.1   & -0.97/0.16\\
UTMOST & 6   & 5.8\pm 1.9   & 1.2\pm 0.4 & 0.3\pm 0.7 & -0.17/0.68  & 1\pm 2         & 0.8\pm 0.7   & -0.15/0.58\\
\enddata
\tablecomments{Uncertainties are the standard deviation of results after bootstrapping 100 times. The nominal power law index from \citet{2014MNRASYoung} is $\beta=1.29$.}
\tablenotetext{a}{The log-likelihood ratio, $R$, and corresponding $p$-values with respect to a power law. 
A negative $R$ corresponds to favoring the opposing model.}
\end{deluxetable*}

\subsection{Single pulses and nulls}\label{sec:sp}
\begin{figure*}
\centering
\includegraphics[width=\textwidth]{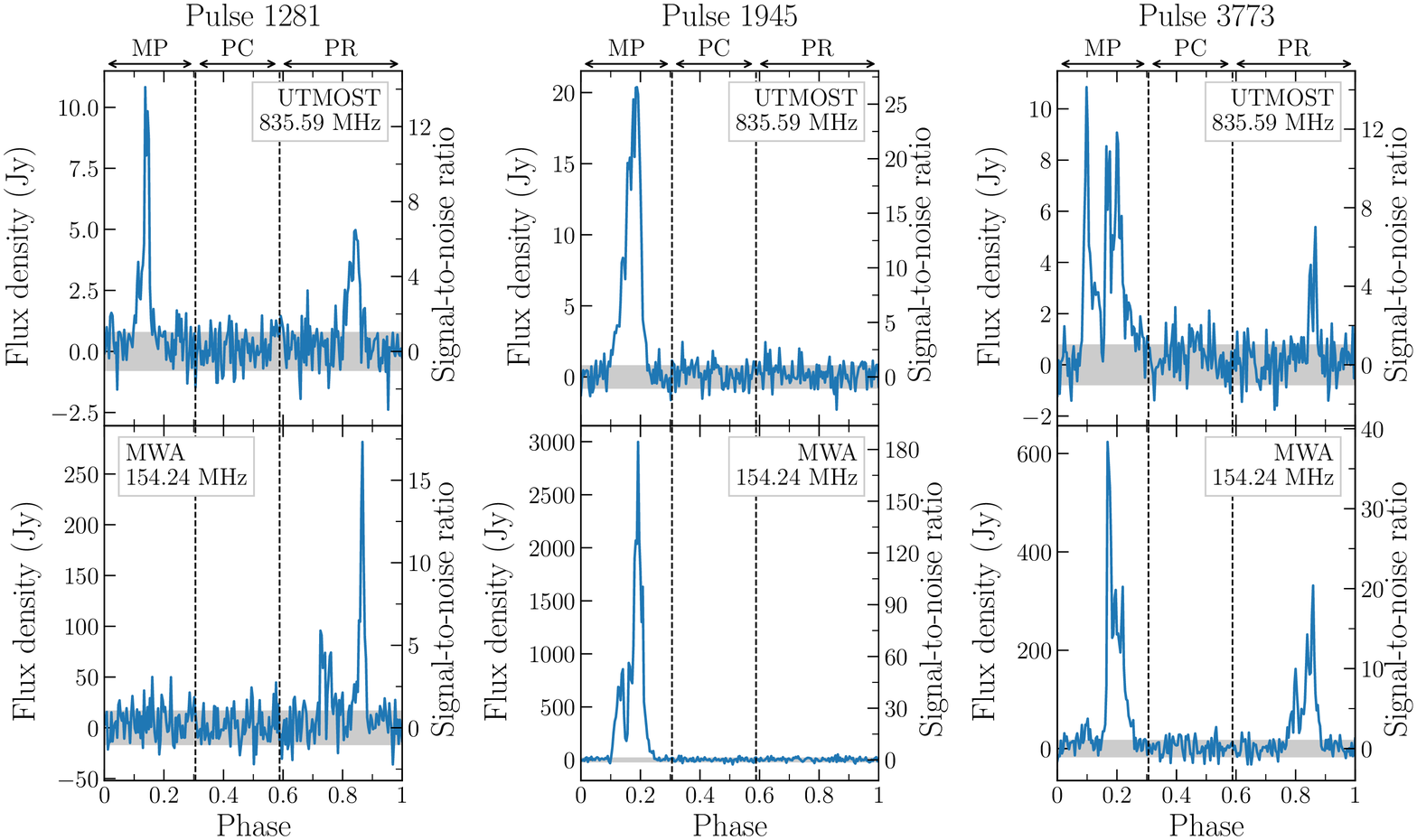}
\caption{Examples of coincident single pulses between UTMOST (top) and the MWA (bottom).
There is a wide variety of pulse shapes and intensities, including instances where entire components are apparently missing at one frequency (e.g. the right-most column of pulses shown here).
These examples also imply that the spectral index is varying from pulse-to-pulse and between components substantially, which is not unexpected.
The gray-shaded region indicates the $\pm1\sigma$ noise level.
Each pulse is titled with its pulse number, corresponding to the number of rotations since the first simultaneous MWA and UTMOST rotation. \label{fig:spgallery}}
\end{figure*}

As with most long-period pulsars, there is wide variation in the pulse-to-pulse structure and intensities from PSR \psr\ (see Figure~\ref{fig:spgallery}).
This behavior is present at both frequencies.
Specifically, counterparts are not always seen for all bright pulses, even to the level of individual pulse components.
There are examples of emission within both the MP and PR locations at one frequency but not necessarily at the other (e.g. the first column in Figure~\ref{fig:spgallery}). 

Within each of the emission regions, the number of pulses detected are also different.
The MWA detected 40 pulses in the MP phase region and 43 pulses in the PR region (a 1:1 ratio), whereas UTMOST detected 241 and 38 in the MP and PR regions, respectively (a 6:1 ratio).
With additional detections, it would be possible to (statistically) probe this aspect of the pulse-to-pulse variation and intermittency.

The median number of rotations between detected pulses is $31\pm 6$ and $9\pm 1$ pulsar rotations for the MWA and UTMOST, respectively.
These could be considered apparent nulls, though in this case a null is simply defined as a $\lesssim 6\sigma$ single pulse event. 
In reality, there are weaker pulses visible in the time series of both telescopes, ao the numbers we present here should be considered only as upper limits.

In addition to the above, it appears that the pulsar enters its bright state earlier at UTMOST frequencies, and also finishes later than emission at MWA frequencies. 
There are four MP pulses, one PC pulse, and one PR pulse detected by UTMOST before the first MWA pulse in the respective phase regions.
The first MWA pulse in any phase window arrives 14 pulsar rotations ($\sim 3.5$\,s) after the first UTMOST pulse.
Additionally, there are three MP pulses and two PR pulses detected by UTMOST after the last MWA pulse in the respective phase windows. 
The last UTMOST pulse from any phase window arrives 53 pulsar rotations ($\sim 13.4$\,s) after the final MWA pulse.

Scaling the peak flux densities of these UTMOST pulses to MWA frequencies, with their corresponding component spectral index (see Section ~\ref{sec:spectral_index}), and comparing to the nominal flux density threshold placed on the MWA single pulse detection process, we expected to see at least five of the 11 UTMOST single pulses (the weaker pulses could feasibly have fallen below the MWA's flux density threshold).
This then suggests that, while the intermittency properties are roughly broadband, the details of the sporadic emission are different at widely spaced frequencies, where in this case we observed a handful of bright single pulses before and after that nominal bright state has begun/ended at the other frequency.

\subsection{Pulse rates and intermittency}\label{sec:rates}
In 5319 rotations of the pulsar, we detected \mwaN\ and \utmostN\ pulses from the MWA and UTMOST, respectively, corresponding to overall pulse rates of $0.06\,\mathrm{s^{-1}}$ and $0.21\,\mathrm{s^{-1}}$.
In general, we see that the number of PC and PR components detected at each telescope is similar, but the number of pulses arriving in the MP window are far fewer at MWA frequencies, regardless of the steep measured spectral index.
For a component-wise split and summary, see Table~\ref{tab:rates}.

\begin{deluxetable}{lCCCCC}
\tablecaption{Pulse rates for each component of PSR \psr\ above the given minimum peak flux density detected, $\min\{S_\nu\}$.\label{tab:rates}}
\tablehead{\colhead{Telescope} & \colhead{$\min\{S_\nu\}$} & \colhead{Total} & \colhead{MP} & \colhead{PC} & \colhead{PR} \\
& ($\rm Jy$)& \colhead{($\rm hr^{-1}$)} & \colhead{($\rm hr^{-1}$)} & \colhead{($\rm hr^{-1}$)} & \colhead{($\rm hr^{-1}$)}}
\startdata
MWA    & 136 & 230 & 107 & 8 & 115\\
UTMOST & 3.3 & 760 & 645 & 13 & 101\\
\enddata
\end{deluxetable}

We conducted nine simultaneous observations over the course of six months (excluding MJD 57865) using the MWA in conjunction with UTMOST and detected PSR \psr\ in the bright state only once. 
While we are limited by small number statistics (in terms of both attempts and detections), this indicates that at MWA frequencies, the duty cycle of bright-state emission from PSR \psr\ is $\delta \sim 11\%$. 
Additional observations (and detections) will be required to further constrain this number, but in general it seems to roughly agree with the few estimates available in the literature, i.e. $\delta \sim 5\text{--}8\%$ \citep{2014MNRASYoung,2016MNRASHobbs}.

\begin{figure*}
\centering
\includegraphics[width=\linewidth]{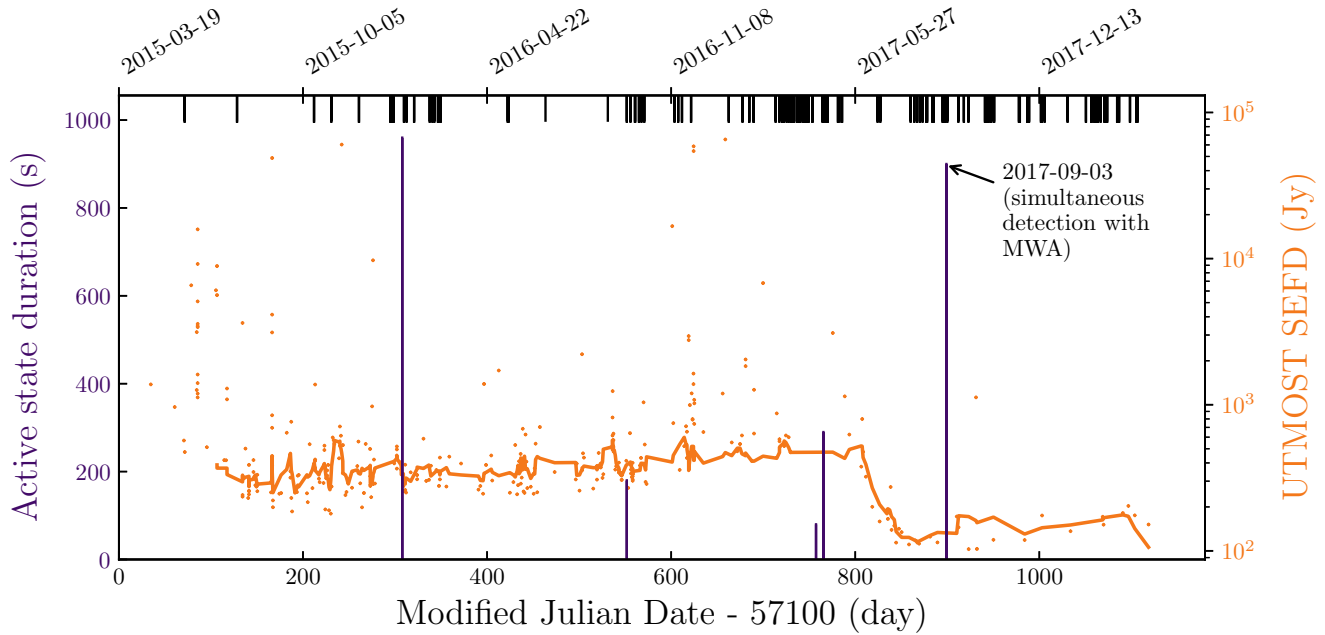}
\caption{UTMOST observations, detections and sensitivity over time. 
Observations made without a bright state detection are shown as black lines at the top of the figure. 
Detections and the respective bright state duration are drawn in dark purple (left ordinate). 
Only for the first and last bright state detections was the full pulse train observed (i.e. beginning and end of the bright state was captured). 
The orange points are the estimated system equivalent flux density (SEFD) for UTMOST over time (right ordinate), calculated by calibrating the system to a standard bright pulsar (PSR J1644--4559).
A 20-point running average, excluding the extreme outliers, is also drawn (solid orange line) to give a more representative measure of the sensitivity over time.
\label{fig:utmost_detections}}
\end{figure*}

In addition to the contemporaneous observations with the MWA, we examined archival UTMOST data taken since mid-2015 for examples of both bright-state and single pulse detections (the ``weak'' state).
Of 148 observations (including those simultaneous with the MWA), ranging in length from $\sim 100\text{--}3600$\,s and spanning $\sim 1200$ days (see Figure~\ref{fig:utmost_detections}), there were only five bright-state detections. 
The sensitivity of UTMOST has varied drastically in time due to hardware maintenance and upgrades, as well as the reconfiguration into a transit-only telescope (starting around 2017 May, lasting $\sim 2$ months).
Thus, the SEFD of the instrument is also included in Figure~\ref{fig:utmost_detections} as a sensitivity indicator.
The bright-state duty cycle from UTMOST detections is $\delta \sim 3.4\%$, which is slightly lower than previously reported.

We can also calculate a duty cycle based on the duration of time PSR \psr\ for which was detected in the bright state with respect to the total observing time (e.g. \citealp{2012MNRASYoung,2014MNRASYoung})\footnote{The methods described by these authors are not applicable in our case, given the low number of detections, the irregularity in observing times and durations, and telescope sensitivity.}.
The bright state duty cycle evaluated in this way is $\delta \sim 2.2\%$, again lower than previous estimates.
This estimate comes with the caveat that, for three of the five detections, the bright-state pulse train was not fully sampled (i.e. observations started after the bright state had already begun, or finished before it ended), so there is a bias toward a smaller duty cycle.

In the remaining observations, there were a total of 19 possible single-pulse detections, but only five of these proved to be significant (one of which is the single pulse we simultaneously detected on 2017 August 5).
The pulsar emits in the weak state much more frequently than the bright state, so we believe that we are sensitivity-limited for both telescopes, and only capture the brightest pulses of the weak-state emission.

\subsection{Rotation measure}\label{sec:rm}
The degree of Faraday rotation the radio emission from a source experiences when traversing the ISM is quantified by the rotation measure (RM).
Typically, the RM is estimated by measuring the change in the polarization position angle across the observed bandwidth (e.g. \citealp{2008MNRASNoutsos,2018ApJSHan}) and thus requires the detector to measure the radiation polarization properties.
While the UTMOST detects only right-hand circular polarization, the MWA is capable of producing fully polarimetric (Stokes $I$, $Q$, $U$ and $V$) data (S. Ord et al. submitted).
Currently, the polarization response is undergoing self-consistency tests (M. Xue et al. submitted) and cross-validation (S. Tremblay et al. in prep.) that will be described in forthcoming publications, hence we do not provide a polarization profile.
However, because RM estimates do not rely on absolute polarimetric calibration, we do provide an estimate of the RM for PSR \psr\ from the MWA data.
Performing RM synthesis \citep{2005A&ABrentjens} on the $6\sigma$ pseudo-integrated profile (Figure~\ref{fig:profile_6sig}) produces $\mathrm{RM_{obs}} = \RMobs \mathrm{\,rad\,m^{-2}}$.
The ionospheric RM contribution was calculated to be $\mathrm{RM_{ion}} \approx \RMion \mathrm{\,rad\,m^{-2}}$ using an updated version of \textsc{ionFR}\footnote{\url{http://ascl.net/1303.022}} \citep{2013A&ASotomayor-Beltran}, so that the latest version of the International Geomagnetic Reference Field (IGRF12; \citealp{2015EP&SThebault}) could be used as an input, along with International GNSS Service vertical total electron content maps (e.g. \citealp{2009JGeodHernandez-Pajares}).
Thus, the ISM contribution is $\mathrm{RM_{ISM}} = \mathrm{RM_{obs}}-\mathrm{RM_{ion}} = \RMism \mathrm{\,rad\,m^{-2}}$.
This is consistent with the previously published value of $23\pm 3 \mathrm{\,rad\,m^{-2}}$ \citep{2014MNRASYoung}, which does not account for ionospheric contributions.
In order to comment further, we would require additional observations of PSR \psr\ in the bright state in order to obtain a better integrated profile and average over ionospheric effects in order to minimize their contribution to the uncertainties.

\section{Low-frequency detection prospects}\label{sec:det_prospects}
With the next generation of radio telescopes on the horizon, there will soon be new opportunities for pulsar searches.
In particular,  precursor instruments such as the MWA can be used to gain valuable insights into what can be expected in terms of survey yields (e.g. \citealp{2017PASAXue}).
Furthermore, studying the frequency evolution of pulsar profiles down to the single pulse level is imperative to correctly measuring the pulse-to-pulse energetics, such as the spectral index.

The intermittency of PSR \psr\ is broadband, in that the pulsar switches between bright and weak states contemporaneously (see also Section~\ref{sec:sp}), over frequencies separated by at least a factor of five.
The bright-state duty cycle is comparable across frequencies, with $\delta\sim 2\text{--}11\%$ corresponding to approximate interburst (or ``off'') timescales of a few hours, and the bright-state duration is between one and 45 minutes.
PSR \psr\ exhibits a relatively rare combination of bright emission, a moderate DM, and relatively active state-switching.
Therefore, this pulsar serves as an interesting link between nulling, RRATs, and state-switching (intermittent) pulsars.
For instance, its interburst timescale is similar to some RRAT burst-rates\footnote{See \url{http://astro.phys.wvu.edu/rratalog/} for published RRAT data}; moreover, it also nulls during its bright emission state.
\citet{2014MNRASYoung} showed that, if the pulsar were a factor of $\sim 4$ farther from Earth, it could be detected as a RRAT when in its weak mode, while the bright state could always be detected in both single-pulses and the average profiles (at 20\,cm).
We cannot comment on the weak-mode emission; however, if we increased the noise in the MWA data by a factor of $\sim 16$, which artificially replicates moving the pulsar to be $\sim 4$ times as distant, then our combined pulse rate drops to $\sim 19\,\mathrm{hr^{-1}}$, similar to many RRATs. 
Furthermore, the steep spectral index and the extent of emission in longitude ($\gtrsim 180^\circ$ based on Figure~\ref{fig:profile_6sig}) argues in favor of the low-frequency detectability of pulsars similar to PSR \psr.

In particular, planned pulsar and fast-transient searches using instruments such as the MWA and SKA-Low are in an advantageous position to find more such sources.
The large field of view and the ability to regularly return to observing fields offer better prospects in detecting (and monitoring) objects like this. 
As an example, detection of super-bright bursts, like that observed with the MWA in this work (Figure~\ref{fig:brightPC}) with a peak flux density of $\sim2.5$\,kJy, imply that conventional single-pulse search and transient search pipelines would be capable of detecting these objects.
While more sensitive telescopes at higher frequencies (e.g. Parkes, MeerKAT) may reveal the subtler features in the underlying emission, particularly in the apparent ``off'' emission states, the initial discoveries and identification of spectrally steep components will likely be made by these low-frequency wide field telescopes (e.g. \citealp{2015MNRASSobey}).

\section{Summary and Conclusions}\label{sec:conclusion}
The simultaneous detection of PSR \psr\ in its bright emission state with the MWA (154.25\,MHz) and UTMOST (835.59\,MHz) on 2017 September 3 marks the first low-frequency detection of this object.
Our detections allowed a measurement of the spectral index, which for the main pulse component is relatively typical of the average pulsar population, where $\alpha_{\rm MP}=\alphaMP$, whilst the so-called precursor component is steeper than average, with $\alpha_{\rm PR}=\alphaPR$.
We also characterized the fluence distributions of two prominent profile components (MP and PR), and contrary to previous results (which indicated a power law distribution), find that a log-normal distribution is the most appropriate fit for both the MWA and UTMOST data.

In addition to properties relating to the emission physics, the MWA also provides an excellent opportunity to study the effects of the ISM.
We measured the DM of PSR \psr\ and found that it required a correction to the cataloged DM ($\DMcat \mathrm{\,pc\,cm^{-3}}$; \citealp{2006MNRASLorimer}) of $\delta{\rm DM}=\DMoffset \mathrm{\,pc\,cm^{-3}}$.
Our improved value ($\rm DM=\DMobs \mathrm{\,pc\,cm^{-3}}$) is consistent with the previous estimate within uncertainties, but is $\sim 50\times$ more precise.
We also used the MWA data to estimate the RM of this pulsar, measuring $\mathrm{RM_{ISM}} = \RMism \mathrm{\,rad\,m^{-2}}$ (after subtracting the ionospheric contribution), which is also consistent with---and about an order of magnitude more precise than---the previously published value ($23\pm 3 \mathrm{\,rad\,m^{-2}}$; \citealp{2014MNRASYoung}).
Further observations of this pulsar, particularly with the MWA, will help to better constrain the DM and RM, in addition to the pulse profiles.

The next generation of wide-field, wide-bandwidth radio telescopes and their pulsar search surveys are on the horizon.
Detection and characterization of objects like PSR \psr\ over a wide frequency range is fundamentally important to developing a comprehensive understanding of the Galactic pulsar population, and more broadly the Galactic neutron star population.

\acknowledgements
B.W.M. and V.G. would like to thank Willem van~Straten for valuable discussion of the single-polarization correction applied to UTMOST flux densities.
The authors acknowledge the contribution of an Australian Government Research Training Program Scholarship in supporting this research.
This scientific work makes use of the Murchison Radio-astronomy Observatory, operated by CSIRO. 
We acknowledge the Wajarri Yamatji people as the traditional owners of the Observatory site. 
Support for the operation of the MWA is provided by the Australian Government (NCRIS), under a contract to Curtin University administered by Astronomy Australia Limited. 
We acknowledge the Pawsey Supercomputing Centre, which is supported by the Western Australian and Australian Governments. 
We acknowledge the Australian Research Council grants CE110001020 (CAASTRO) and the Laureate Fellowship FL150100148.
Part of this research was supported by the Australian Research Council Centre of Excellence for All Sky Astrophysics in 3 Dimensions (ASTRO 3D), through project number CE170100013.
This paper made use of archived data obtained through the Australia Telescope Online Archive and the CSIRO Data Access Portal (http://data.csiro.au).

\facilities{MWA, Molonglo Observatory}
\software{Astropy (\citealp{2013A&AAstropy,2018AJAstropy}; v2.0.2), DSPSR \citep{2011PASAvanStraten}, ionFR (\citealp{2013A&ASotomayor-Beltran}), Matplotlib (\citealp{Hunter:2007}; v2.0.2), powerlaw (\citealp{2014PLoSOAlstott}; v1.4.1), PSRCHIVE \citep{2004PASAHotan,2012AR&TvanStraten}, STILTS \citep{2006ASPCTaylor}}

\appendix
\section{Correcting UTMOST flux density measurements of a polarized source}\label{sec:appendix}
For the complex electric field vector $\varepsilon$, the total intensity (Stokes $I$) can be described as $I = \varepsilon_{\rm L}^2 + \varepsilon_{\rm R}^2$, where $\varepsilon_{\rm L}$ and $\varepsilon_{\rm R}$ are the left- and right-hand circularly polarized voltages induced in orthogonally polarized receptors.
It is clear then that a detector sensitive to only one polarization will detect $50\%$ of the total intensity from an unpolarized source (where $\varepsilon_{\rm L}$ and $\varepsilon_{\rm R}$ are equal but completely out of phase).

The measured SEFD for UTMOST is calibrated to nominally unpolarized pulsars, so the quoted SEFD of 120\,Jy implicitly corrects for the fact that the instrument only detects half of the total flux density.
In the case where the target source is polarized, we must also correct for the polarized component of the total flux density.
PSR \psr\ is slightly circularly polarized with $V/I=-0.1$, which indicates that $\varepsilon_{\rm R}^2 > \varepsilon_{\rm L}^2$ (by the PSR/IEEE standard, Stokes $V$ is positive for left-hand circularly polarized radiation; see \citet{2010PASAvanStraten}).
UTMOST is therefore detecting $50\%$ of the unpolarized flux density (i.e. $45\%$ of the total power) and $100\%$ of the polarized flux density (i.e. $10\%$ of the total power).
Due to the SEFD implicitly correcting by a factor of 2 (i.e. accounting for the fact that only $50\%$ of the unpolarized flux density is detected), the measured UTMOST flux densities for PSR \psr\ will be overestimated (i.e. $2\times (45\% + 10\%) = 110\%$ of the total flux density).
We therefore must correct the UTMOST flux densities by a factor of $\zeta=1/1.1\approx 0.9$.

\bibliographystyle{aasjournal}
\bibliography{references}



\end{document}